\documentclass[aps,pre,bibtex,onecolumn,notitlepage,superscriptaddress]{revtex4-1}

\usepackage{graphicx}
\usepackage{dcolumn}
\usepackage{bm}
\usepackage{color}
\usepackage{hyperref}
\usepackage{amsmath}
\usepackage[capitalise]{cleveref}
\usepackage{amsfonts}
\usepackage{amsthm}
\usepackage{cases}
\usepackage{amssymb}
\usepackage{latexsym}
\usepackage{multirow}
\usepackage{xcolor}
\usepackage{mathrsfs}
\usepackage{textcomp}
\usepackage{lipsum}
\usepackage[normalem]{ulem} 
\hypersetup{colorlinks=true,linkcolor=blue,citecolor=blue}

\newcommand{\ve}[1][K]{\mathbf{#1}}

\begin{document}

\title{Correlations between rare events  for Gaussian stochastic processes with  long-term memory}

\author{Apurba Biswas}
\affiliation{Univ. Bordeaux, CNRS, LOMA, UMR 5798, F-33400, Talence, France.}
\author{Thomas Gu\'{e}rin}
\email{thomas.guerin@u-bordeaux.fr}
\affiliation{Univ. Bordeaux, CNRS, LOMA, UMR 5798, F-33400, Talence, France.}

\begin{abstract}
Rare events refer to  qualitatively unlikely events whose realization can nevertheless have important consequences. 
Typically, the prediction of the kinetics of these events relies on Arrhenius laws, with exponentially distributed waiting times, and no correlations between  successive occurrences. 
However, this description breaks down in the presence of long-term memory, as has been observed in the contexts of geophysical time series or protein dynamics. 
So far, existing analytical approaches do not quantify the correlations between rare events due to long-term memory. 
Here, for non-Markovian Gaussian processes, we determine analytically the impact of long-term memory on the distribution of first and second passage times to a rarely reached threshold,  using a perturbation approach. 
This distribution is non-exponential, thus going beyond the Arrhenius paradigm. 
 We obtain an explicit expression for the covariance between the first and second passage times, and we predict how the mean time to the next extreme event depends on the previous passage time, illustrating the phenomenon of clustering of extreme events.  These analytical results, validated through extensive stochastic simulations, shed lights on the strong correlation between successive occurrences of extreme events due to long-term memory. 
\end{abstract}

\maketitle

Many physical  and chemical processes are controlled by ``rare'' events,  referring to  qualitatively unlikely events that can nevertheless have important consequences when they are realized \cite{hanggi1990reaction,pollak2005reaction,chowdhury2022extreme}. Such events appear in various areas of science, as exemplified by the formation or rupture of   bonds \cite{hanggi1990reaction,bullerjahn2014theory,bullerjahn2020non,bullerjahn2016analytical,jeppesen2001impact} in chemical physics, molecular motor dynamics~\cite{Badoual2002a,guerin2011}, nucleation events, stock market crashes~\cite{Bouchaud1998}, and climate~\cite{ragone2018computation} or population~\cite{kamenev2008colored,dykman2008disease} dynamics.  

A key question about rare events is to characterize how often they occur.   
The paradigm describing this occurrence kinetics is given by Arrhenius (or Kramers) laws, in which (i) the distribution of times to reach a rare event  (waiting times) is  exponential and does not depend on initial conditions, and  
(ii) the mean waiting time is exponentially large with the energy barrier that has to be overcome to reach the target configuration \cite{hanggi1990reaction} (or the pseudo-potential barrier for non-equilibrium systems~\cite{Freidlin1984,Maier1992,bouchet2016generalisation,delacruz2018minimum}). These Arrhenius laws are derived in the weak noise limit by analyzing the dynamics at the top of the (pseudo-)potential barrier.  
In this limit,  waiting times usually become  larger than all relaxation times of the dynamics, and are thus independent of initial conditions. Also, for the same reason, there are no correlations between successive occurrences of rare events. 

However, these properties may be questioned for stochastic processes $x(t)$ with  long-term memory, for which correlation functions decay as a power-law rather than exponentially, \textit{i.e.}, which satisfy  
\begin{equation}
 \phi(\tau)\equiv \lim_{t\to\infty}\frac{\langle x(t)x(t+\tau)\rangle}{\langle x^2(t)\rangle }\underset{\tau\to\infty}{\simeq} \frac{A}{\tau^{\alpha}} \label{llm},
\end{equation}
where $A\neq 0$, $0<\alpha<1$ and $\langle x(t)\rangle=0$ by convention. The above property, hereafter referred to as ``long-term memory property''~\cite{santhanam2008return,bunde2005long}, typically arises for processes resulting from the evolution of an infinite number of degrees of freedom with broadly distributed timescales, for example, in the context of the dynamics of polymers~\cite{Panja2010}, proteins~\cite{kou2004generalized,Min2005,granek2005fractons,yang2003protein}, interfaces \cite{ReviewBray}, critical Ising models \cite{zhong2018generalized},  active baths \cite{sarkar2025tracer}, neurons~\cite{middleton2003firing},  earthquakes \cite{lennartz2008long} or rainfalls \cite{bunde2013there}. For long-term memory processes,  the relaxation time to reach a stationary state is infinite, and thus cannot be smaller than the Arrhenius time. This questions the validity of the oversight of initial conditions and of the independence between successive occurrences of extreme events. Even though for Gaussian processes with long-term memory, mathematical results indicate that Arrhenius laws are still valid
\cite{pickands1969upcrossing,pickands1969asymptotic}, deviations from the Arrhenius paradigm are observed in different contexts. First, the phenomenon of \textit{clustering of extreme events} has been identified and is relevant for various geophysical time series (rainfalls, temperatures...)\cite{bunde2005long,carpenter2022long}. Second, long-term memory may be involved in the non-exponential distribution of barrier crossing times \cite{min2006kramers,goychuk2009viscoelastic} which are commonly observed in protein dynamics \cite{karplus2000aspects, sunney2002single,lu1998single,flomenbom2005stretched,english2006ever}.

Although the effect of memory on rare event kinetics has been 
investigated recently~\cite{ferrer2021fluid,ginot2022barrier,lavacchi2020barrier,lavacchi2022non,bullerjahn2020non,kappler2018memory,Caraglio2018,carlon2018effect,medina2018transition,goychuk2007anomalous,Sliusarenko2010,arutkin2020extreme,levernier2020kinetics,delorme2017pickands,goswami2023effects,zanovello2021target,ayaz2021non}, most analytical approaches overlook long-term memory. Existing theories including long-term memory rely on the use of generalized Fokker-Planck equation \cite{goychuk2007anomalous} (which is inconsistent with simulations \cite{singh2019comment,bullerjahn2017unified}), 
the Wilemski-Fixman~\cite{Sliusarenko2010} or independent return intervals~\cite{santhanam2008return} approximations (which neglect a number of memory effects), including phenomenological parameters to describe temporal fluctuations of energy barriers (dynamic disorder) \cite{zwanzig1990rate,goychuk2009viscoelastic}, or approximately dividing between slow and fast variables \cite{bullerjahn2020non}. Recently, an approach that analyzed the paths after the rare event was proposed but predicted only the mean waiting times \cite{Barbier2024}. Hence, the analytical prediction of their distribution, as well as of the correlations between successive rare events, remains unresolved. 

\begin{figure}[t!]
 \centering
\includegraphics[width=8cm]{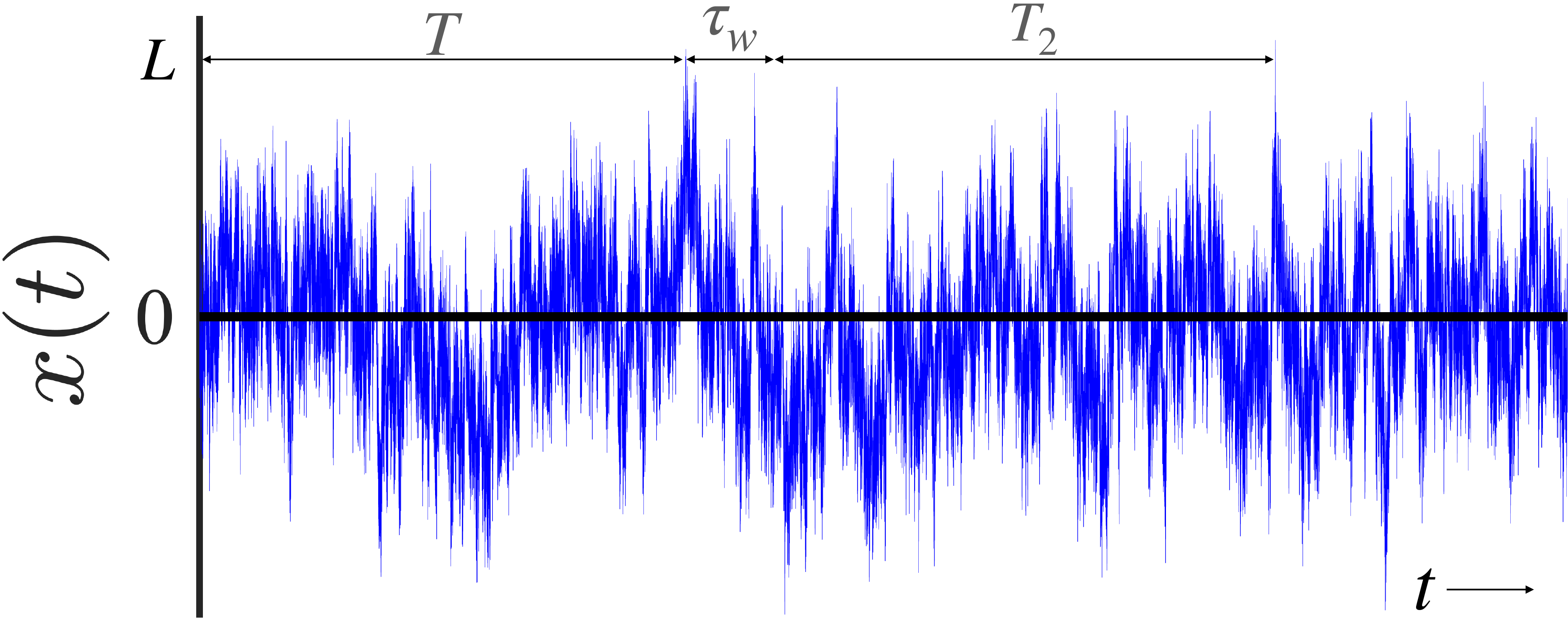}
\caption{\textbf{Sketch of the problem}. 
For a non-smoooth Gaussian stochastic process $x(t)$ with long-term memory, what is the distribution of the first passage time $T$ to a rarely reached threshold $L$ ? More generally,  
what are the correlations between  $T$ and the second passage time, $T_2$ (measured after a lag-time $\tau_w$) ? The shown trajectory corresponds to a process $x(t)$ satisfying Eq.~(\ref{GLE}), with $\alpha=1/2$.} 
\label{FigSketch}
\end{figure}

Here, we address this question in the case of Gaussian processes with long-term memory (Fig.~\ref{FigSketch}). We develop a theory that analytically predicts  how the distribution of first passage times to a rarely reached threshold deviates from an exponential distribution, due to long-term memory. 
The theory is then extended to obtain simple and explicit formulas describing the correlations between the first and second passage times, in excellent agreement with extensive simulations, identifying the key parameters controlling the dependence of the future rare event times on past ones. 
 
\textit{Model. } We first consider the position $x(t)$ of an overdamped particle in a harmonic potential (with stiffness $k$), submitted to  non-instantaneous friction   with kernel $K(t)$, at temperature $\mathcal{T}$,  obeying the Generalized Langevin Equation (GLE) 
\begin{align}
 &\int_{0}^t dt' K(t-t')\dot{x}(t')=-k\  x(t)+\xi(t),  \label{GLE}
\end{align}
where the centered Gaussian noise $\xi(t)$ is correlated as $ \langle \xi(t)\xi(t')\rangle= k_B\mathcal{T}K(\vert t-t'\vert)$. We focus on scale invariant kernels $K(t)= K_{\alpha}/[\Gamma(1-\alpha)t^{\alpha}]$, so that $x(t)$ satisfies the long-term memory property   (\ref{llm}) with $ A=K_{\alpha}/[\Gamma(1-\alpha)k]$. With these hypotheses, $x(t)$ describes a number of physical processes, such as a tracer particle in a viscoelastic fluid \cite{mason1995optical,gisler1998tracer,mason1997particle}, a tagged monomer of a long   polymer \cite{Panja2010,Panja2010a,bullerjahn2011monomer}, or the  distance between two protein residues as experimentally observed \cite{Min2005}. Our aim is to characterize the statistics of the first passage time (FPT), called $T$, of $x(t)$ to a threshold value $L$ that is anomalously high ($L\gg l$, with $l=\langle x^2\rangle_{s}^{1/2}=\sqrt{k_B\mathcal{T}/k}$). Reaching $L$ requires to overcome the energy barrier $\Delta E = kL^2/2  \gg k_B\mathcal{T}$ and is thus  rare. 
We will also characterize the second passage time $T_2$, defined as the time to obtain a second threshold crossing.  Although we consider here a passive model, all our results will be generalized to  active ones (see below). 
    
Since Kramers escape times usually depend on the dynamics near the top of the potential barrier, here identified as $x=L$, we first identify the characteristic timescale $t^*$ and length scales $l^*$ of this dynamics.
From dimensional analysis, we identify $l^*=k_B\mathcal{T}/(kL)$, where $kL$ is the slope of the potential at the target position $L$. At short times, the fluctuations of $x(t)$ read 
$\text{var}(x(t))_{x(0)=L}\simeq \kappa t^{2H}$ with $H=\alpha/2$ and $\kappa=2k_B\mathcal{T}/[\Gamma(1+\alpha)K_\alpha]$. This implies that $x(t)$ is non-smooth, \textit{i.e.,} displaying non-differentiable trajectories (like Brownian motion) \cite{ReviewBray}. Defining $t^*$ as the time at which the spatial fluctuations of $x(t)$ become of the order of $l^*$ leads to  $t^*=(l^*/\sqrt{\kappa})^{1/H}$. 

 \textit{First passage to a threshold: deviation from Arrhenius law.- } 
Our starting point is the following generalized renewal equation \cite{guerin2016mean}
\begin{align}
p(L,t)=\int_0^t d\tau \ F(t-\tau) \ p(L,t\vert  T=t-\tau) \label{RenewalEq},
\end{align}
where $t>0$, $p(x,t)$ is the probability density function (PDF) of $x(t)$, $F(t)$ 
is the  PDF of first passage times, and $p(L,t\vert T=t')$ is the probability density of observing $x(t)=L$ given that  the FPT is $t'$, if the process is allowed to continue after the first passage.  We assume that $x$ starts with a stationary distribution, so that  $p(x,t)=p_s(x)=e^{-kx^2/(2k_B\mathcal{T})}/\sqrt{2\pi l^2}$. 
The exact Eq.~(\ref{RenewalEq}) is obtained by partitioning the probability to observe $x(t)=L$  
over the values of the FPT, which is necessarily smaller than $t$  due to the non-smooth property of trajectories.  

Our approach consists of exploiting the fact that the first passage kinetics results from the contribution of two widely separated timescales. The first relevant timescale $t^*$ characterizes  the dynamics near the threshold. 
A second important timescale is the mean FPT in the weak noise limit, called $T_{\mathrm{RE}}$, which grows as $T_{\mathrm{RE}}\propto e^{\Delta E/k_B\mathcal{T}}$ \cite{Barbier2024} and is thus much larger than $t^*$. Since $F(t)$ varies at the scale $T_{\mathrm{RE}}$, it is essential to understand how the propagator behaves at this timescale. A key remark is that, following an event where $x(0)=L$ (not for the first time), the relaxation to the steady state is algebraic
\begin{align}
p(L,t\vert  L,0) \underset{t\to\infty}{\simeq} p_s(L) \left(1+ \frac{L^2 A}{l^2 t^\alpha} \right) \simeq p_s(L) \left(1+    \frac{\varepsilon}{\overline{t}^\alpha} \right) . 
\end{align} 
This expression can be derived from the known solutions of Eq.~(\ref{GLE}). In the second equality, we have used the rescaled time $\overline{t}=t/T_{\mathrm{RE}}$, and this has led us to introduce  the parameter $\varepsilon\equiv A L^2/(l^2T_{\mathrm{RE}}^\alpha)$ which will be a key small parameter of our theory. 
Our strategy is then to expand the relevant quantities in powers of $\varepsilon$: we set $F(t)\simeq T_{\mathrm{RE}}^{-1}  [F_{0}(\overline{t})+\varepsilon F_1 (\overline{t} ) + ...  ] $, while for the propagator after the FPT, we use 
\begin{align}
p(L,t  \vert T=t- \tau)\simeq p_s(L)\left[1+\varepsilon\  R\left(\overline{\tau},\overline{t}-\overline{\tau} \right)+...\right], \label{DefR}
\end{align}
with $R$ a dimensionless function and $\overline{\tau}=\tau/T_{\mathrm{RE}}$. In turn, if the time $\tau$ after the FPT is of the order of $t^*$, dimensional analysis indicates that  
 $p(L,t  \vert T=t-\tau)\simeq  q^*(\tau/t^*)/l^*$, where $q^*$ is a dimensionless scaling function. 
 
Next, we decompose the integral over $\tau$ in Eq.~(\ref{RenewalEq})  into two parts: one for $\tau\sim  t^*$ (where the propagator is proportional to $q^*$), and another part for $\tau \sim  T_{\mathrm{RE}}$ [where we use Eq.~(\ref{DefR})]. Using this procedure, we obtain
\begin{align}
&p_s(L)=\frac{F_{0}(\overline{t})+\varepsilon F_1(\overline{t})}{l^*T_{\mathrm{RE}}}t^*    c_H+ p_s(L)
 \int_0^{\overline{t}} d\overline{\tau}  [F_0(\overline{t}-\overline{\tau})+\varepsilon F_1(\overline{t}-\overline{\tau})] [1+\varepsilon R\left( \overline{\tau},\overline{t}-\overline{\tau}\right) ],\label{9421}
\end{align}
where $c_H=\int_0^\infty dv \ q^*(v)$ is a numerical constant depending only on $H$. 
For $\varepsilon=0$, the above integral equation for $F_0$ that can be readily solved, yielding
\begin{align}
&F_0(\overline{t})= e^{-\overline{t}},& T_{\mathrm{RE}}=\frac{t^*  c_H}{l^* p_s(L)}. \label{Order0Pred}
\end{align}
As expected, we obtain the exponential distribution of the Arrhenius paradigm, with a mean time that is exponentially large with the energy barrier, $T_{\mathrm{RE}}\propto 1/p_s(L)\propto e^{\Delta E/k_B\mathcal{T}}$. 
The above formula is consistent with the mathematical results of Pickands \cite{pickands1969asymptotic,pickands1969upcrossing} and previous results \cite{levernier2020kinetics}.    

To obtain the effects of long-term memory, we analyze (\ref{9421}) at order $\varepsilon$, obtaining a relation between $R$ and $F_1$:
\begin{align}
F_1(\overline{t}) + \int^{\overline{t}}_0 d\overline{\tau}  \left[ F_1(\overline{t}-\overline{\tau}) + e^{-\overline{t}+\overline{\tau}} R  (\overline{\tau}, \overline{t}-\overline{\tau}  ) \right] =0. \label{eqn for h and R}
\end{align}
To obtain an equation for $R$, we write a generalized version of the renewal equation (\ref{RenewalEq}), for any $t>0$, and any set $0<t_1<t_2<....<t_m$ and $x_1,x_2,...,x_m$:
\begin{align}
&p(L,t ;x_1, t+t_1;...;x_m, t+t_m) = 
 \int_0^t dt'  F(t') p(L,t;x_1,t+t_1;...;x_m,t+t_m\vert T=t'). \label{renewal eqn_nPoints}
\end{align}
To close the system, we assume that the  trajectories after the first passage $T=t$ follow Gaussian statistics, with a covariance equal to the covariance of trajectories conditional to $x(t)=L$ (not for the first time). The function $R$ is in fact the rescaled mean of the trajectories in the future of the FPT, and an equation for $R$ can be found by investigating the above equation for $m=1$ with the above mentioned method of taking into account all timescales and collecting terms of order $\varepsilon$ (see  end matter). The resulting equations for $R$ and $F_1$ can be explicitly solved, leading to
\begin{align}
F_1(\overline{t})= \overline{t}^{1-\alpha} \frac{\overline{t} -2+ \alpha}{(2-\alpha)(1-\alpha)} e^{-\overline{t}}. \label{Result_h}
\end{align}
This formula provides the first corrections of $F$ to the exponential behavior due to long-term memory, which take the form of a universal function depending only on $\alpha$. 
It was obtained in a perturbation approach based on the separation of timescales $t^*\ll T_{\mathrm{RE}}$ in the rare event limit,  assuming Gaussian statistics after the first passage time $t=T$ with a covariance approximated by the covariance of initial trajectories  conditional to $x(t)=L$. This kind of approximations was found to be very accurate  in other contexts~\cite{guerin2016mean,levernier2020kinetics}. Here they can be self-consistently justified by noting that, at order $\varepsilon$, equation (\ref{renewal eqn_nPoints}) is satisfied for all $m$ and all $x_i$ as soon as it is satisfied for $m=1$, suggesting that they are in fact exact at order $\varepsilon$ at the timescale $T_{\mathrm{RE}}$ (see SM \cite{citeSM}). 
Using the above expressions, 
one obtains for the moments $\langle T^n\rangle=T^{n}_{\mathrm{RE}}  \left[n! +  \varepsilon n\Gamma(n+2-\alpha)/ (2-3\alpha+\alpha^2)\right]$, which is compatible with the results of Ref.~\cite{Barbier2024} (restricted to $n=1$). 

\begin{figure}[t!]
 \centering
\includegraphics[width=12cm]{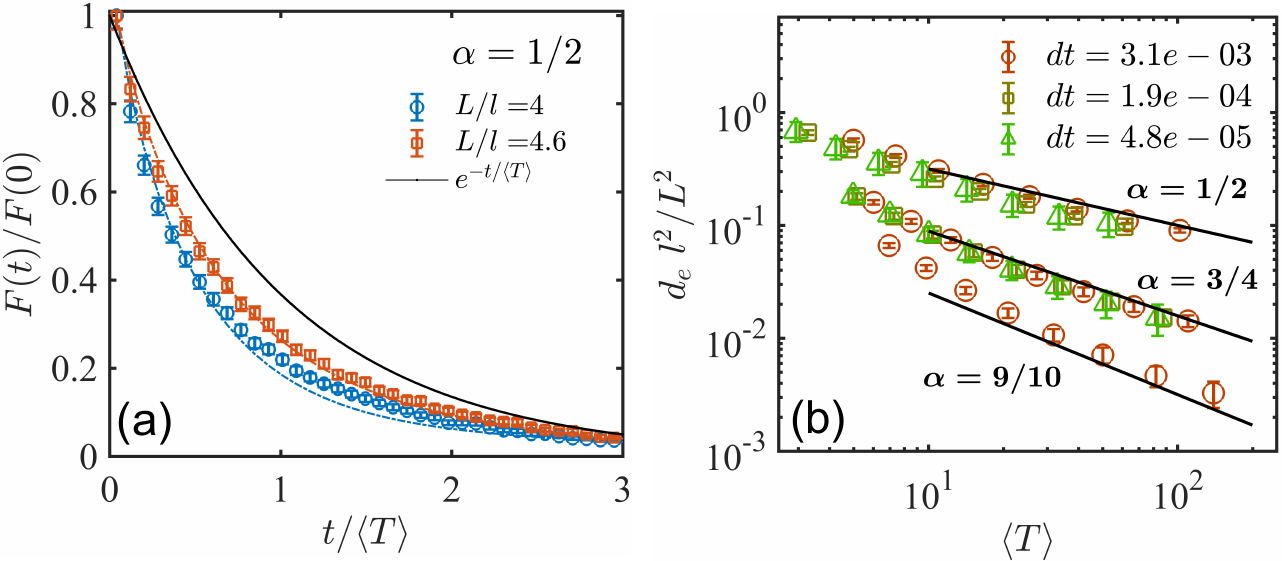}
\caption{ (a) The distribution $F(t)$ of FPTs, renormalized by $F(0)$ for better readability, is plotted 
for the processes described by Eq.~(\ref{GLE}) for $\alpha=1/2$. Symbols: simulations. Dashed lines:  prediction 
$F(t)/F(0)=e^{-v}\left\{1+  \varepsilon \left[v^{1-\alpha}\left(-2+\alpha+v\right)-\Gamma(3-\alpha)v  \right]/[(2-\alpha)(1-\alpha)] \right\}$ with $v=t/\langle T\rangle$ and $\varepsilon=AL^2/(l^2\langle T\rangle^{\alpha})$, which is compatible at order $\varepsilon$ with Eq.~(\ref{Result_h}). 
(b)  Quantity $d_el^2/L^2$, measuring the deviation of $F(t)$ to the exponential distribution, for the same process, with $\alpha=1/2,  3/4$ and $9/10$. Symbols: simulations, with the value of the used time step in legend. Lines:   prediction (\ref{Predict_de}). Units were chosen so that $K_\alpha=k=k_B\mathcal{T}=1$. }
\label{FigFPTDistribution}
\end{figure} 

To test our approach,  we have generated stochastic trajectories $x(t)$ satisfying Eq.~(\ref{GLE}) with stationary initial conditions by using a modified version of the circulant matrix algorithm \cite{davies1987tests,dietrich_fast_1997}, as done in Ref.~\cite{Barbier2024}. Trajectories are  sampled exactly at times $t_n=n\ dt$, with $dt$ the time step,  and we measured the distribution of FPT, see Fig.~\ref{FigFPTDistribution}(a). Deviations from exponential distributions are clearly observed, and correctly captured by our theory, at least when they are small (as expected for a perturbation calculation). To further quantify the deviation from exponential behavior, we define     $d_e =\ \langle T^2\rangle/\langle T\rangle^2-2$,
so that $d_e=0$ for exponentially distributed $T$. In our theory, 
\begin{align}
d_e     
\simeq 2 \Gamma(2 - \alpha) \frac{L^2 A}{l^2 \langle T\rangle^\alpha}.\label{Predict_de}
\end{align}
This formula is validated numerically for various values of $\alpha$ in Fig.~\ref{FigFPTDistribution}(b), indicating that our theory is able to quantify the first corrections to exponential behavior of the FPT distribution due to long-term memory.

\begin{figure}[t!]
 \centering
\includegraphics[width=12cm]{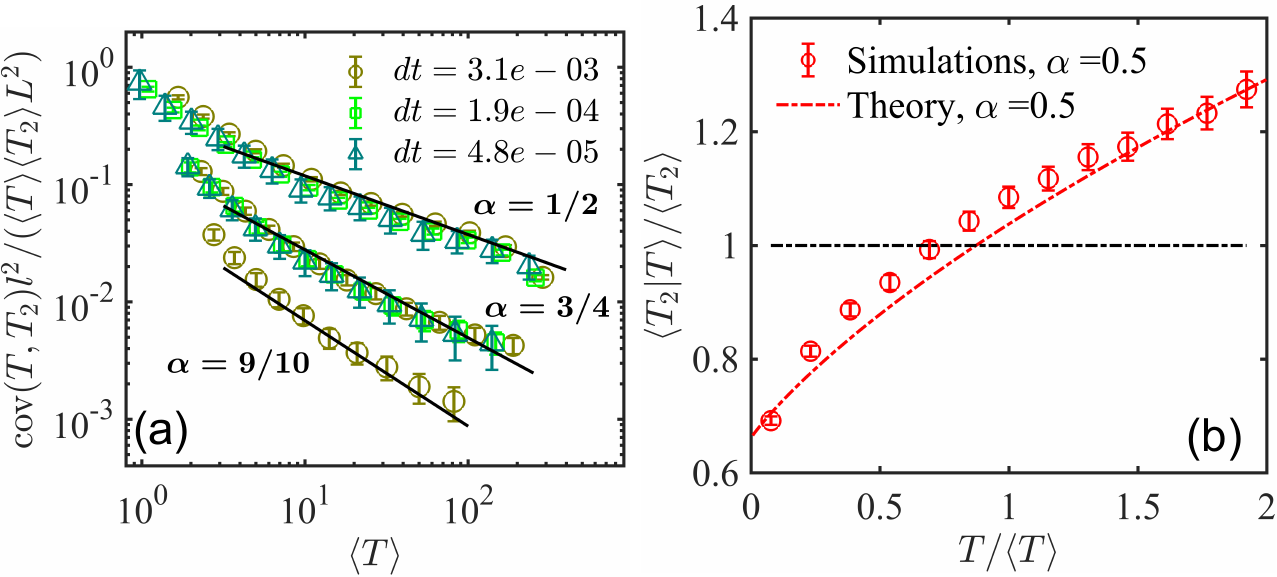}
\caption{  (a) Normalized covariance between the first and second passage times $T$ and $T_2$ for the passive process described by the GLE  (\ref{GLE}). Symbols: simulations.  Lines:  analytical prediction (\ref{CovT1T2}). Here, $\tau_w=1$. 
(b) Mean  second passage time $T_2$, conditional on $T$, against $T/\langle T\rangle$ for the same process.  
 Symbols: simulation results. Line: prediction (\ref{T1CondT2}), with $\varepsilon=L^2A/(l^2\langle T\rangle^{\alpha})$. Parameters: $\tau_w=1$, $L/l=3.6$, $dt=1.91\times 10^{-4}$. Units are such that $K_\alpha=k=k_B\mathcal{T}=1$.  
}\label{Fig_Cov_T1_T2}
\end{figure}

\textit{Correlations between successive  passage times. }
We now ask the question of the second passage time $T_2$.  Actually, many threshold crossings are expected at times $t\sim t^*$ after the first passage, and for a non-smooth process the next-passage time is formally zero. Instead of looking at these short-times recrossing events, we focus on the more relevant quantity, which is the time to come back to the threshold once the particle has moved away from it. Hence, we  introduce a lag-time $\tau_w$ and we define $T_2$  so that, after $T+\tau_w$, the threshold is hit at $T+\tau_w+T_2$ for the first time, and $T_2$ does not depend on  $\tau_w$ provided $t^*\ll \tau_w\ll T_{\mathrm{RE}}$ (see SM  \cite{citeSM}). 

To predict the statistics of $T_2$ conditional to a fixed value of $T$, we see $T_2$ as a first passage problem for the process in the future of the real first passage time, for which we know the mean and the covariance as outputs from the previous analysis. We thus adapt the same procedure to take into account these complicated initial conditions, leading to an analytical formula of the conditional distribution of $T_2$ given $T$. Since we know the distribution of $T$, this leads to determining the full joint distribution of $T_2,T$ (see  End Matter).  In particular, for their rescaled covariance, we find
\begin{align}
\frac{\text{cov}(T,T_2) }{\langle T\rangle \langle T_2\rangle}\simeq \frac{A L^2}{2 l^2  \langle T \rangle^{\alpha}} \Gamma(3-\alpha). \label{CovT1T2}
\end{align}
This is the main result of the present work. Of note, although our theory is based on a perturbative expansions in powers of $\varepsilon$, here the perturbation results for the covariance between $T,T_2$ is actually the leading order term because in the Arrhenius paradigm there are no correlations between successive rare events.  It is validated by comparing with simulations for various values of $\alpha$, see Fig.~\ref{Fig_Cov_T1_T2}(a). Equation (\ref{CovT1T2}) indicates that $T$ and $T_2$ are positively correlated. Physically, this can be understood as follows: if $T$ is unusually large, it is likely that, due to the other degrees of freedom, the energy barrier to reach the threshold is higher than usual, and one can expect that the next passage will tend to be higher than usual because it also faces this higher energy barrier which, due to the long-term memory, will not relax fast to its average value.  
Our theory is of interest as it provides the information about the correlations between successive events without assuming \textit{a priori} a dynamics for this barrier, 
leading to a simple law depending only on a restricted number of parameters: $A$ (appearing in the large time behavior of the correlation function), the threshold $L$, the mean time $\langle T\rangle $ and the stationary amplitude of fluctuations $l^2=\langle x^2(t)\rangle_s$.

 \begin{figure}  
\includegraphics[width=12cm]{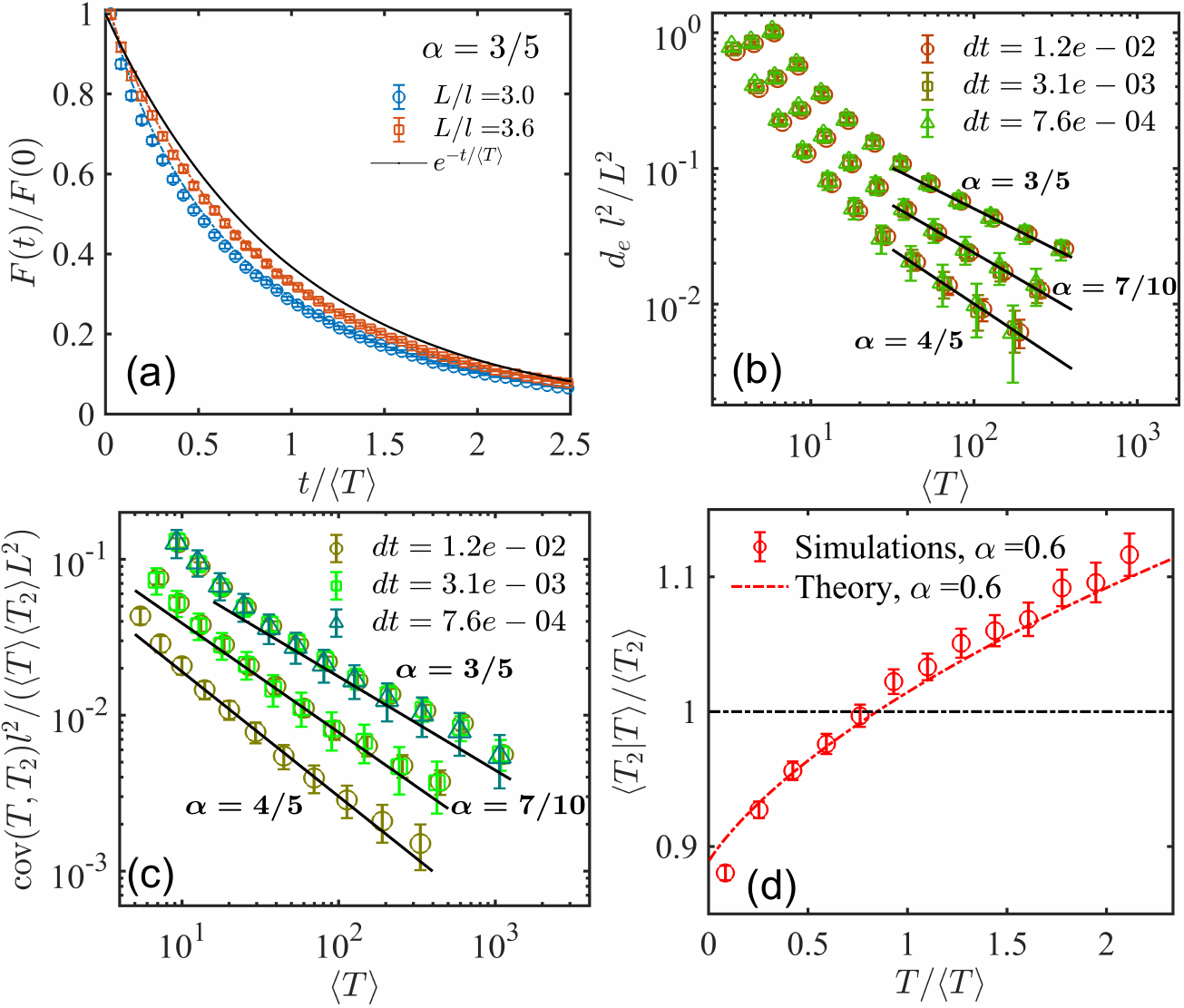}
\caption{Same quantities as in Figs \ref{FigFPTDistribution} and \ref{Fig_Cov_T1_T2}, but for the active process described by Eq.~(\ref{ActiveProcess}). 
(a) Distribution of FPTs, $F(t)$. 
(b) Parameter $d_e$ quantifying the deviation to the exponential distribution.
(c) Rescaled covariance between $T$ and $T_2$. 
(d) Mean second passage conditioned on the first passage time.  
 Symbols: simulations. Lines:   predictions. 
Parameters: $\tau_w=1$ [(c),(d)] and 
 $L/l=2.8$, $dt=1.2\times 10^{-2}$ [(d)]. \label{Fig_Active} }
\end{figure} 

To further illustrate the correlations between $T,T_2$, we consider the expected time to the second passage given the first passage, for which we obtain
\begin{align}
\frac{\langle T_2\vert T = \tau \rangle}{\langle T_2 \rangle} \simeq 1 + \varepsilon\        \frac{e^{\bar{\tau}} \Gamma(2-\alpha,\bar{\tau})-\Gamma(3-\alpha) }{1-\alpha} \label{T1CondT2},
\end{align}  
where $\overline{\tau}=\tau/\langle T\rangle$. This expression  is in good agreement with numerical simulations, see Fig.~\ref{Fig_Cov_T1_T2}(b).  It also shows that the time  to the next event can be expected to be higher than its global average if the previous passage time was observed to be unusually large. 
The positive correlations between successive events [Eqs.~(\ref{CovT1T2}) and (\ref{T1CondT2})], 
together with the non-exponential distribution of waiting times [Eq.~(\ref{Result_h})] are characteristic features of  the clustering of extreme events~\cite{bunde2005long} that our theory quantifies analytically in the rare event limit  for Gaussian processes. 

\textit{Generalization to active processes.} Our theory is not restricted to passive processes 
but applies to any centered  non-smooth stationary Gaussian process with long-term memory. 
Indeed, our results (\ref{Order0Pred}), (\ref{Result_h}), (\ref{CovT1T2}), (\ref{T1CondT2}) hold when one identifies $l^2=\langle x^2\rangle_s$, while 
 $l^*=l^2/L$ and $t^*=(l^*/\sqrt{\kappa})^{1/H}$, where $\kappa$ and $H$ are defined such that $\text{var}(x(t))_{x(0)=L}\simeq \kappa t^{2H}$ at short times. As an example of active process, we consider   $x(t)$ satisfying  
 \begin{align}
& \dot{x}=-x + \xi(t), &\langle \xi(t+\tau)\xi(t)\rangle=\frac{2H(2H-1)}{\tau^{2-2H}}, \label{ActiveProcess}
 \end{align}
 that displays long-term memory with $\alpha=2-2H$~\cite{Sliusarenko2010}.  The values of $\kappa,l,A$ for this process are given in SM  \cite{citeSM}.  Figure \ref{Fig_Active} shows that our formulas for the distribution of FPT and the coupling between the first and second passage times are in good agreement with simulations for this active process. 
 
 \textit{Conclusion.- } We have introduced a theoretical analysis showcasing  the analytical link between the long-term memory properties and the clustering of extreme events \cite{bunde2005long}. 
 On the basis of non-Markovian Gaussian stochastic processes with long-term memory, we have quantified the deviations between the distribution of first passage times to a rarely reached threshold to an exponential distribution, thus going beyond the Arrhenius paradigm. 
The theory also accounts for the joint statistics of first and second passage times, whose correlations are characterized by a small number of parameters.  
These parameters depend on the dynamics either close to the target or near the stable position, both of which are expected to be locally Gaussian even for non-Gaussian dynamics. We may thus expect that our theory could be generalized to non-Gaussian processes with long-term memory. In the SM  \cite{citeSM}, for viscoelastic subdiffusion in an anharmonic potential (which is non-Gaussian), we show that correlations between successive passages exist and are well described by our formulas.   It would also be interesting to see if our theory could quantify such correlations for other models with slowly decaying memory of initial conditions, such as self-interacting processes \cite{azencott2018large,coghi2025accelerated}.   

Our method can be iterated to calculate the statistics of  
the $n^{\text{th}}$ passage time $T_n$, and the correlations between $T_n$ and $T_{m}$. Our results could thus have application to the barrier crossing problem with viscoelastic memory, where each passage in the vicinity of the barrier can lead to a barrier crossing with a low probability.  
Our results are validated using extensive numerical simulations for both active and passive stochastic processes with long-term memory that are relevant in various physical areas (dynamics of polymers, proteins, tracers in viscoelastic fluids...).  Our findings should also be relevant for geophysical time series which often display the long-term memory property  \cite{bunde2005long,bunde2013there}.
Altogether, our results shed light, with analytical methods, on the correlations between successive occurrences of rare events due to long-term memory,  beyond Arrhenius laws.
  
   \begin{acknowledgments}
A.B. and T.G. acknowledge support of the grant \textit{ComplexEncounters}, ANR-21-CE30-0020. T.G. thanks Olivier Bénichou, Raphael Voituriez and Hamid Kellay for interesting discussions. Computer time for this study was provided by the computing facilities MCIA (Mesocentre de Calcul Intensif Aquitain) of the Universit\'e de Bordeaux and of the  Universit\'e de Pau et des Pays de l’Adour.
\end{acknowledgments}

The data that support the findings of this article are openly available \cite{DataLongMem}. 

\appendix

\begin{center}
\textbf{End matter}
\end{center}

\section{Details on the equations for the statistics of first passage times}

Here, we outline the key steps that lead to the self-consistent equation that determines the function $R$.  
We start by considering Eq.~(\ref{renewal eqn_nPoints}) for $m=1$, multiply this equation by $x_1$, and integrate the resulting equation with respect to $x_1$. This leads to
\begin{align}
&p(L,t)  \mathbb{E} (x(t+t_1) \vert x(t)=L) = \int_0^t du  \ F(t-u) p(L,t  \vert T=t-u)   \mathbb{E}(x(t+t_1)\vert T=t-u;x(t)=L) , \label{EqCond}
\end{align}
where $\mathbb{E}(y\vert B;C)$ represent the conditional average of $y$ given that the conditions $B$ and $C$ are realized. In particular, using existing expressions for conditional  Gaussian variables \cite{Eaton1983}, we have
\begin{align}
&\mathbb{E}(x(t+t_1)\vert x(t)=L)=L \frac{\sigma(t+t_1,t)}{\sigma(t,t)}= L\ \phi(t_1),\label{AvTrajStat}
\end{align}
where $\sigma(t,t')=\text{cov}(x(t),x(t'))$ denotes the covariance of the original process, and we have used $\sigma(t,t')=l^2\phi(\vert t-t'\vert)$ since the stochastic process $x(t)$ is stationary. 

As mentioned in the main text, we evaluate the terms in Eq. (\ref{EqCond}) with the assumption that the process in the future of the FPT is Gaussian, with a mean $\mu(t\vert \tau)\equiv\mathbb{E}(x(t)\vert T=\tau)$, and a covariance $\text{cov}(x(t),x(t')\vert T=\tau) \equiv  \sigma(t,t'\vert \tau)$ that is approximated as the covariance of the initial process given that $x(\tau)=L$ (not necessarily for the first time). 
Applying existing formulas for conditional Gaussian distributions  \cite{Eaton1983}, we can write
\begin{align}
\sigma(t,t'\vert \tau)=&\sigma(t,t')-\frac{\sigma(t,\tau)\sigma(t',\tau)}{\sigma(\tau,\tau)} 
= l^2\left[\phi(\vert t-t'\vert )-\phi(\vert \tau-t\vert)\phi(\vert \tau-t'\vert)  \right] \label{CondSigma},\\
p(L,t  \vert  T=&t-u)=\frac{e^{-[L-\mu(t\vert t-u)]^2/[2\sigma(t,t\vert t-u)]}}{\sqrt{2\pi \sigma(t,t\vert t-u)}} \label{Propag},\\
  \mathbb{E}(x(t+t_1)&\vert T=t-u;x(t)=L) =\mu(t+t_1\vert t-u) 
 -[\mu(t\vert t-u)-L]\frac{\sigma(t+t_1,t\vert t-u)}{\sigma(t ,t\vert t-u)}.\label{CondMeans}
\end{align}
Equations (\ref{EqCond}) and  (\ref{RenewalEq}), together with (\ref{Propag}) and (\ref{CondMeans}), form a  set of coupled integral equations that define $\mu$ and $F$.

 We now analyze these equations in the rare event limit $L\to\infty$. 
First we consider the length and time scales, $l^*$ and $t^*$ respectively when trajectories are near to the threshold at $x=L$. For $t$ of the order of $T_\text{RE}$, and at times $u\sim t^*$ after the FPT, we postulate that the displacements occur at the lengthscale $l^*$, so that
\begin{align}
&\mu(t\vert t- u)\simeq L-l^* f(U), &U=u/t^*, \label{muAtScaletStar}
\end{align}
where we have assumed that the trajectory after the FPT does not depend on the value of the FPT. Then, the PDF of $x=L$ at a time $u\sim t^*$ after the FPT is  
\begin{align}
&p(L,t  \vert T=t-u)\simeq  \frac{q^*(u/t^*)}{l^*} , &q^*(U)=\frac{e^{-\frac{f^2(U)}{2U^{2H}}}}{\sqrt{2\pi} U^{H}}.  \label{PropagqStar}
\end{align}
 We refer to Ref.~\cite{levernier2020kinetics} for the derivation of an equation for $f$, which is involved in the calculation of the mean FPT (in the absence of long term memory). 

Next, for larger times, when $\overline{t}=t/T_{\text{RE}}$ and $\overline{u}=u/T_{\text{RE}}$ are of order $1$, we use the ansatz
\begin{align}
&\mu(t\vert t-u)\simeq \frac{L A}{T_{\text{RE}}^{\alpha}} R\left( \overline{u} , \overline{t}-\overline{u} \right), & (t,u=\mathcal{O}(T_{\text{RE}})) \label{muAtScaleTRE},
\end{align}
which is compatible with Eq. (\ref{DefR}) [see Eq. (\ref{Propag}) with $\sigma(t,t\vert t-u)\simeq l^2 +\mathcal{O}(\varepsilon^2 l^4/(L^4\overline{t}^{2\alpha})\simeq l^2$]. 
Now, we  estimate the terms appearing in Eq.~(\ref{EqCond}) at different time scales, when $t,t_1 \sim \mathcal{O}(T_{\mathrm{RE}})$. First, when $u\sim t^*$,  evaluating (\ref{CondMeans}) with the scaling forms (\ref{muAtScaletStar}) and (\ref{muAtScaleTRE}) for $\mu$ and the fact that $u\ll t_1$,  we obtain 
\begin{align}
\mathbb{E}&(x(t+t_1)  \vert T=t-u;x(t)=L) \simeq \frac{LA}{T^{\alpha}_{\mathrm{RE}}} R\left( \overline{t_1}, \overline{t}\right)  +l^* f\left(\frac{u}{t^*}\right)    \frac{\phi(t_1)-  \phi(t_1) \phi(u)}{1-\phi^2(u)} \simeq \frac{LA}{T^{\alpha}_{\mathrm{RE}}} R\left( \overline{t_1}, \overline{t}\right),
\label{CondMeanSmallu}
\end{align} 
where in the second equality we have used that $\phi(u)\simeq1$, $\phi(t_1)\simeq A/t_1^{\alpha}$, and   $l^*\ll L$.
 Next, we evaluate the same quantity for $u \sim \mathcal{O}(T_{\mathrm{RE}})$, still with $t,t_1=\mathcal{O}(T_{\mathrm{RE}})$ which leads to 
 \begin{align}
&\mathbb{E}(x(t+t_1)\vert T=t-u;x(t)=L) \simeq \frac{LA}{T^{\alpha}_{\mathrm{RE}}} R\left( \overline{t_1}+\overline{u}, \overline{t}-\overline{u} \right) - \left( \frac{LA}{T^{\alpha}_{\mathrm{RE}}} R\left( \overline{u}, \overline{t}-\overline{u} \right)  -L\right)   \phi(t_1)  \nonumber \\
&\simeq \frac{LA}{T^{\alpha}_{\mathrm{RE}}} \left[R\left( \overline{t_1}+\overline{u}, \overline{t}-\overline{u} \right)  +  \frac{1}{\overline{t_1}^\alpha}\right], \label{CondMeanLargeu}
\end{align}
where in the last equality we have used $\phi(t_1)\simeq A/t_1^\alpha$. 

We can now evaluate the integral in Eq. (\ref{EqCond}), where we take into account a contribution with $u=\mathcal{O}(t^*)$ [where we use (\ref{CondMeanSmallu}) and (\ref{PropagqStar})] and the contribution of $  u=\mathcal{O}(T_\text{RE})$ [where we use (\ref{CondMeanLargeu}) and $p(L,t\vert T=t-u)\simeq p_s(L)$], we obtain at lowest order  in $\varepsilon$: 
\begin{align}
 \frac{1}{\overline{t}^{\alpha}_1}& = e^{-\overline{t}} R \left( \overline{t}_1, \overline{t} \right)  
+ \int^{\overline{t}}_0 d\overline{u}   \left(R \left( \overline{t}_1+ \overline{u}, \overline{t}-\overline{u} \right)+ \frac{1}{\overline{t}^{\alpha}_1}\right)e^{\overline{u}- \overline{t}   }. 
 \label{self consistent eqn for R}
\end{align}
This self-consistent equation for $R$ and   can be considerably simplified if one introduces $W(\overline{t}_1,y)=e^{-\overline{t}}R(\overline{t}_1,t)$, with $y=\overline{t}+\overline{t}_1$, and $v=\overline{u}+\overline{t}_1$ (inside the integral), leading to
\begin{align}
W(\overline{t}_1,y)+\int_{\overline{t}_1}^{y}dv \ W(v,y)=\frac{e^{-y+\overline{t}_1}}{\overline{t}_1^\alpha}.
\end{align}
This Volterra integral equations of the second kind for $W(\cdot,y)
$ admits the known solution \cite{HandbookIntegralEquations}
\begin{align}
W(\overline{t}_1,y)=\frac{e^{-y+\overline{t}_1}}{\overline{t}_1^\alpha}+\int_y^{\overline{t}_1}dv \ e^{\overline{t}_1-v} \frac{e^{-y+v}}{v^\alpha}.
\end{align}
Simplifying the integral and expressing the result for the quantity $R$  leads to
\begin{align}
R(\overline{t},\overline{u}) =\frac{  \overline{t}^{-\alpha} \left( 1-\alpha+\ \overline{t}\right)-  (\bar{t}+\bar{u})^{1-\alpha}}{1-\alpha}. \label{sol of R}
\end{align}
With this value of $R$, the equation (\ref{eqn for h and R}) for $F_1$ can be readily solved upon using the Laplace transform, leading to the final solution \eqref{Result_h}.

\section{Details on the equations for the statistics of second passage times}

We define $F_2(\tau_2\vert \tau)$ as the PDF of the second passage   time (SPT) $T_2$ conditional to $T=\tau$. 
Let us define the stochastic process after the FPT, given that $T=\tau$, as $y(t \vert\tau)\equiv x( t-(\tau_w+\tau))$, for all trajectories with $T=\tau$. Our strategy is to apply the methods of the previous section to this process, which is still Gaussian, but non-centered. In particular, at the scale $T_{\mathrm{RE}}$, its mean reads
\begin{align}
\langle y(t\vert\tau)\rangle \simeq \frac{LA}{T_{\mathrm{RE}}^\alpha}R(\overline{t}+\overline{\tau_w},\overline{\tau})\simeq \frac{LA}{T_{\mathrm{RE}}^\alpha}R(\overline{t},\overline{\tau}),
\end{align}
where $\overline{\tau}_w=\tau_w/T_{\mathrm{RE}}\ll 1$. For its covariance, one has $\text{cov}[y(t\vert\tau), y(t'\vert\tau)]\simeq \sigma(t+\tau,t'+\tau\vert\tau)$. Importantly, at the scale $t,t'=\mathcal{O}(T_{\mathrm{RE}})$, $\text{cov}[y(t\vert\tau), y(t'\vert\tau)]\simeq l^2\phi(\vert t-t'\vert)$ is the same as that of the original process $x(t)$. 

Next, let us note $p_2(y,t)$ the PDF of $y(t\vert \tau)$ (here we omit the fixed quantity $\tau$ in $p_2$).  Using the above expressions for the mean and variance of $y$ and formulas for conditional Gaussian distributions \cite{Eaton1983}, we obtain
\begin{align}
&p_2(L,t ) \simeq p_s(L) \left(1+\varepsilon R(\overline{t}, \overline{\tau})  \right),\label{BehavP2}\\
&\mathbb{E} (y(t+t_1\vert \tau) \vert y(t\vert \tau)=L) \simeq \frac{LA}{T^{\alpha}_{\mathrm{RE}}}\left( R\left(\bar{t}+\bar{t}_1, \bar{\tau} \right)+ \frac{1}{\bar{t}^\alpha_1} \right). \label{tOrderTreEq1}
\end{align}

We anticipate that  the distribution of SPTs is exponential in the absence of long-term memory, so that we look for solutions under the form
\begin{align}
&F_2(\tau_2\vert  {\tau})=\frac{1}{T_{\mathrm{RE}}}\left[e^{-\bar{\tau}_2}+\varepsilon \ h_2\left(\bar{\tau}_2\vert \bar{\tau}\right)+...\right],
\end{align}
where $\bar{\tau}_2={\tau}_2/T_{\mathrm{RE}}$. We also work with the assumption that the trajectories after the second passage, given that the second passage is $\tau_2$, are Gaussian distributed, with a covariance approximated by the covariance of $y$ conditional to $y(\tau_2\vert\tau)=L$, and a mean which, at the timescale $T_{\mathrm{RE}}$, behaves as
\begin{align}
\mathbb{E}(y(t \vert \tau)\vert T_2=t-u ) = \frac{LA}{T_{\mathrm{RE}}^\alpha}R_2(\overline{u},\overline{t}-\overline{u}\vert\overline{\tau}). 
\end{align}
Also, we assume that, at the timescale $t^*$, the statistics of paths after the second passage does not depend on the value of the second passage, and it is the same as the one after  the first passage. As a consequence,  we can use the same equations (\ref{RenewalEq}) and (\ref{EqCond}) as for the analysis of $x(t)$, if we replace $F$ by $F_2$, $p$ by $p_2$, $x$ by $y$ and $R$ by $R_2$. The only differences comes from the evaluation of $p_2(L,t)$ and the conditional value of $\langle y(t+t_1 )\rangle$ which are given by (\ref{BehavP2}) and (\ref{tOrderTreEq1}). Applying the procedure of the previous section we finally obtain
\begin{align}
 R(\bar{t}&+ \bar{t}_1 , \bar{\tau}  ) + \frac{1}{\bar{t}^\alpha_1} = e^{-\bar{t}}R_2\left(  \bar{t}_1 , \bar{t} \vert \bar{\tau} \right)  +\int_0^{\bar{t}}d\bar{u} \ e^{-(\bar{t}-\bar{u})}\left[R_2\left(  \bar{t}_1+\bar{u}, \bar{t}-\bar{u} \vert \bar{\tau} \right) +   \frac{1}{\bar{t}_1^\alpha} \right], \label{EqR2}\\
R(\overline{t},&\overline{\tau})= h_2(\overline{t} \vert \bar{\tau}) + \int^{\overline{t}}_0 d\overline{u} \ h_2(\overline{t}-\overline{u} \vert \bar{\tau}) + \int^{\overline{t}}_0 d\overline{u} \ e^{-(\overline{t}-\overline{u})} R_2 \left(\overline{u}, \overline{t}-\overline{u} \vert \bar{\tau} \right) . \label{eqn for h and R2}
\end{align}
Note that these equations for $R_2,h_2$ are the same as the equations for $R,F_1$ [compare with Eqs (\ref{eqn for h and R}) and (\ref{self consistent eqn for R})], except for additional terms in the left-hand sides, which come  from Eqs. (\ref{BehavP2}) and (\ref{tOrderTreEq1}). These equations can be solved by standard methods described in the SM  \cite{citeSM}, leading to explicit values for $R_2$ and $h_2$ and then, since one knows the distribution $F$, to an explicit expression for the joint distribution of $T,T_2$.

\begin{center}
\textbf{SUPPLEMENTAL MATERIAL}
\end{center}
 
 Here, we provide calculation details supporting the results shown in the main text, including:
\begin{itemize}
\item a description of the stochastic processes studied in this work (Section \ref{SecStocProc}),
\item a detailed solution to several equations for the calculation of first passage and second passage time distributions   (Section \ref{FPTDistr}),
\item a proof through illustration that the result for the  normalized covariance is independent of the lag time $\tau_w$, (Section~\ref{lag time dependence}), 
\item a generalization to one example of non-Gaussian process (Section \ref{non-gaussian process}),
\item an argument showing that some of the  approximations of the theory are  exact (Section \ref{SectionRelaxApprox}).
\end{itemize}

\section{Details on the stochastic processes under study}
\label{SecStocProc}

In the main text, we considered passive and active models  described by Eqs.~(\ref{GLE}) and (\ref{ActiveProcess}) respectively. Their correlation functions $\phi$ can respectively be found in Refs.~\cite{goychuk2007anomalous} and \cite{Sliusarenko2010}. For the sake of completeness, here we provide the function $\phi$ and its properties. For  the passive process, in 
units chosen so that $K_\alpha=k=k_B\mathcal{T}=1$ (as used in our simulations),  one has $l=\langle x^2\rangle_{s}^{1/2}=\sqrt{k_B\mathcal{T}/k}=1$ and 
\begin{align}
&\phi(\tau)=E_\alpha\left(-\tau^\alpha \right),&A=\frac{1}{ \Gamma(1-\alpha) }, \ \ \ &\kappa=\frac{2 }{\Gamma(1+\alpha)}, \ \ \ \ \ \ \ \ H=\frac{\alpha}{2},
\end{align}
where $E_\alpha$ is the Mittag-Leffler function. Next, for the active process, one has~\cite{Sliusarenko2010} 
\begin{align}
&l^2=\langle x^2\rangle_s=\Gamma(1+2H), \ \ \ \ \alpha=2-2H, \ \ \ \ \ \ A=\frac{2H(2H-1)}{\Gamma(1+2H)}, \\
&\phi(\tau)= \frac{ e^{-\tau} \Gamma(1+2H) + e^{\tau} \Gamma(1+2H,\tau) + \frac{\tau^{1+2H} e^{-\tau}}{1+2H} M(1+2H; 2+2H, \tau) - 2\tau^{2H}}{2 \Gamma(1+2H)},
\end{align}
where $\Gamma(\cdot,\cdot)$ is the  upper  incomplete gamma function and $M$ denotes the Kummer function.  
Note that  long-term memory  exists when $0<\alpha<1$,  corresponding to  $1/2<H<1$, for which $A>0$. 

\section{Details on the solutions to the equations (\ref{EqR2}) and (\ref{eqn for h and R2}) for the FPT and SPT distributions} \label{FPTDistr}

Here we give the solutions to Eqs (\ref{EqR2}) and (\ref{eqn for h and R2}). Equation  (\ref{EqR2}) for $R_2$ is  the same as the equation for $R$, see Eq.~(\ref{self consistent eqn for R}), except for a new term on the left hand side (namely, $R\left(\bar{t}+\bar{t}_1, \bar{\tau} \right)$). Applying the principle of superposition of solutions,  we find that 
\begin{align}
R_2\left(\bar{t}_1 ,\bar{t} \vert \bar{\tau} \right) =R \left(\bar{t}_1 ,\bar{t} \right) + R\left(\bar{t}+\bar{t}_1, \bar{\tau} \right)\label{SOlR2}
\end{align}
is the solution of Eq.~(\ref{EqR2}). Inserting this expression into Eq.~(\ref{eqn for h and R2}), we obtain
 \begin{align}
h_2(\overline{t}\vert \bar{\tau}) + \int^{\overline{t}}_0 d\overline{u} \ h_2(\overline{t}-\overline{u} \vert \bar{\tau}) + \int^{\overline{t}}_0 d\overline{u} \ e^{-(\overline{t}-\overline{u})} R \left(\overline{u}, \overline{t}-\overline{u} \right) =R(\overline{t},\overline{\tau})e^{-\overline{t}}. \label{eqn for h and R2bis}
\end{align}
The above equation for $h_2$ is almost the same as the equation (\ref{eqn for h and R}) for $F_1$, with an additional term on the right-hand side. Its solution can therefore be written as 
\begin{align}
h_2(\bar{t}\vert \bar{\tau})= F_1(\bar{t}) + h^*_2(\bar{t}\vert \bar{\tau}), \label{linearity}
\end{align}
where $h_2^*$ satisfies
 \begin{align}
h^*_2(\overline{t} \vert \bar{\tau}) + \int^{\overline{t}}_0 d\overline{u} \ h^*_2(\overline{t}-\overline{u} \vert \bar{\tau})=  R\left(\bar{t}, \bar{\tau} \right) e^{-\bar{t}}.  \label{eqn for h2star and R}
 \end{align}
Using the expression of $R$ [see Eq.~(\ref{sol of R})], we find that the solution of this equation is
\begin{align}
 h^*_2(\bar{t} \vert \bar{\tau})&=\frac{e^{-\bar{t}}}{(2-\alpha) (1-\alpha) \bar{t}^{\alpha} \bar{\tau}^{\alpha} (\bar{t}+\bar{\tau})^{\alpha}} \Bigg[-\bar{t}^{1+\alpha} \bar{\tau}^{\alpha} (2-\alpha-2\bar{\tau}) - \bar{\tau}^{1+\alpha} \bar{t}^{\alpha} (2-\alpha-\bar{\tau})\nonumber \\
 & - \bar{t}^{\alpha} \bar{\tau}^2 (\bar{t}+\bar{\tau})^{\alpha} + (2-\alpha) (1-\alpha) \bar{\tau}^{\alpha} (\bar{t}+\bar{\tau})^{\alpha} + \bar{t}^2 \bar{\tau}^{\alpha} (\bar{t}^{\alpha} - (\bar{t}+\bar{\tau})^{\alpha}) \Bigg].  \label{sol of hstar} 
 \end{align}
 This solution can be checked by taking the Laplace transform with respect to $t$, and comparing with the Laplace transform of (\ref{eqn for h2star and R}). Finally, the joint  distribution $F_2(\tau,\tau_2)$ of $(T,T_2)$ reads
\begin{align}
F_2(\tau,\tau_2)=F(\tau)F_2(\tau_2\vert \tau)=\frac{1}{T_{\mathrm{RE}}^2}\left(e^{-\bar{\tau}-\bar{\tau}_2}+\varepsilon h_2(\bar{\tau},\bar{\tau}_2) \right), \label{SolFSP}
\end{align}
with $h_2(\bar{\tau},\bar{\tau}_2)=e^{-\bar{\tau}_2}F_1(\overline{\tau})+e^{-\bar{\tau}}[F_1(\overline{\tau}_2)+h_2^*(\bar{\tau}_2\vert\tau)]$.
Using  previous expressions for $F_1$ and $h_2^*$, we finally obtain
\begin{align}
h_2(\bar{\tau},\bar{\tau}_2)=\frac{e^{-\bar{\tau}-\bar{\tau}_2}  \left(\bar{\tau}^{\alpha} \bar{\tau}_2^{\alpha} (\bar{\tau}+\bar{\tau}_2) (\alpha+\bar{\tau}+\bar{\tau}_2-2)+(\alpha-2) (\bar{\tau}+\bar{\tau}_2)^{\alpha} \left(\bar{\tau}^{\alpha} (\alpha+\bar{\tau}_2-1)+\bar{\tau} \bar{\tau}_2^{\alpha}\right)\right)}{\bar{\tau}^{\alpha} \bar{\tau}_2^{\alpha} (2-\alpha) (1-\alpha)(\bar{\tau}+\bar{\tau}_2)^{\alpha}}.\label{h2}
\end{align}
Using Eqs.~(\ref{SolFSP}) and (\ref{h2}), it is straightforward to find the expressions for the  covariance of $T,T_2$ and the conditional mean $\langle T_2\vert T\rangle$ given in the main text. 

\section{Dependence on lag time $\tau_w$\label{lag time dependence}}
\begin{figure}[h!]
 \centering
\includegraphics[width=7cm]{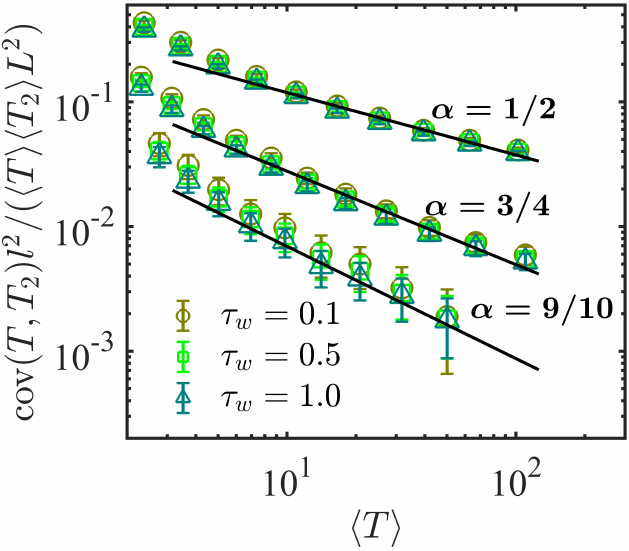}
\caption{Normalized covariance between first and second passage times for the process described by the GLE (\ref{GLE}) for various  $\tau_w$. Symbols: simulations.  Lines:  analytical prediction (\ref{CovT1T2}). Here, $dt=3.1\times 10^{-3}$ and units were chosen so that $K_\alpha=k=k_B\mathcal{T}=1$.  
}\label{Fig_Cov_T1_T2-DEPENDENCE_ON_TAUW}
\end{figure}

In the main text, we claim that the results for the covariance between the first and second passage times, namely $T$ and $T_2$ are independent of the lag time, $\tau_w$, after the first passage, when $t^*\ll \tau_w\ll\langle T\rangle$. A numerical validation for this claim is shown in Fig.~\ref{Fig_Cov_T1_T2-DEPENDENCE_ON_TAUW}.

\section{Generalization to non-Gaussian processes\label{non-gaussian process}}
\begin{figure}[t!]
 \centering
\includegraphics[width=\linewidth]{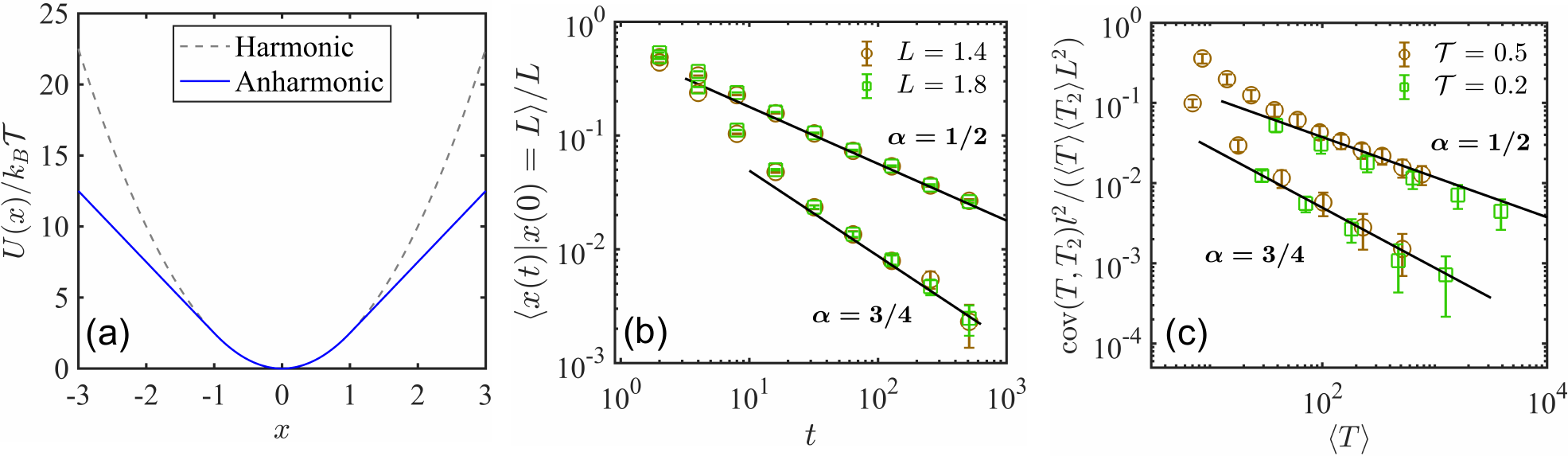}
\caption{(a) Anharmonic potential $U(x)$ scaled by $k_B\mathcal{T}=0.2$ (blue line). The continuation of its harmonic part is also shown for comparison (dashed black line). 
(b) Symbols:   mean of $x(t)$ conditional to $x(0)=L$ at temperature $\mathcal{T}=0.2$, obtained in simulations. Black lines: analytical prediction  (\ref{conditional mean of non-gaussian process}). 
(c) Normalized covariance between the first and second passage times $T$ and $T_2$ versus $\langle T\rangle$ for various choices of $\mathcal{T}$ as indicated in the figure. The choices for the target thresholds are $L\in [1,3]$ for $\mathcal{T}=0.5$ and $L\in [1,1.8]$ for $\mathcal{T}=0.2$. Symbols: simulations.  Lines:  analytical prediction [Eq.~(\ref{CovT1T2}) of main text]. Units were chosen so that $K_\alpha=k_B=k=1$.  
}\label{Fig_anharmonic}
\end{figure}

In this section, we extend our results to an example of non-Gaussian processes. We focus on the physically relevant example of diffusion in an anharmonic potential $U(x)$ described by the GLE  
\begin{align}
 &\int_{0}^t dt' K(t-t')\dot{x}(t')=-U'(x)+\xi(t),&  \langle \xi(t)\xi(t')\rangle= k_B\mathcal{T} K(\vert t-t'\vert ), \ \ \ \ K(t)=\frac{K_\alpha}{\Gamma(1-\alpha)t^\alpha} \label{GLENG}.
\end{align}
As an example of  an anharmonic potential, we choose  [see Fig.~\ref{Fig_anharmonic}(a)]
\begin{align}
U(x)=
\begin{cases}    
    \frac{1}{2}kx^2,         &       \vert x\vert \leq  x_m, \\
    \frac{1}{2}k x^2_m+  k x_m\vert x-x_m\vert , & \vert x\vert \geq  x_m.
\end{cases} \label{non-gaussian potential}
\end{align}
Here, we choose the units so that $K_\alpha=k=x_m=k_B=1$. 
In simulations, we employ the Markovian embedding technique \cite{goychuk2009viscoelastic} to generate the non-Gaussian process described by Eq.~(\ref{GLENG}), in which one approximates the kernel as a sum of $N$ exponential terms, 
\begin{align}
K(t)\simeq \sum^N_{i=1} \frac{K_{\alpha}}{\Gamma(1-\alpha)} C_{\alpha}(b) \left(\frac{\nu_0}{b^i}\right)^\alpha \exp\left(-\frac{\nu_0 t}{b^i}\right), \label{memory kernel expansion}
\end{align}
where $b$, $C_{\alpha}(b)$, $\nu_0$ are parameters of the method~\cite{goychuk2009viscoelastic}. For this class of kernels the noise can be described  as a sum of $N$ independent Ornstein-Uhlenbeck noise components,  see \cite{goychuk2009viscoelastic} for details.  We work with two values of the exponent $\alpha=1/2$ and $\alpha=3/4$. The parameters used are: $N=16,~ \nu_0=10^3,~ b=10,~C_{1/2}(10)=1.3$ and $C_{3/4}(10)=1.85$, for which the exact kernel and its approximated value differ by less than $5\%$ over the range  $t=10^{-3}$ to $10^{10}$. 


Of note, although the potential $U(x)$ is non-linear, at long times, if the temperature is small enough, we can write $\langle U'(x(t))\rangle\simeq U''(0)\langle x(t)\rangle=\langle x(t)\rangle$, and the long-time limit of the averaged GLE becomes
\begin{align}
\langle x(t)\rangle_{x(0)=x_0}= - \int_0^tdt' \frac{1}{\Gamma(1-\alpha)(t-t')^\alpha} \dot{x}(t')\underset{t\to\infty}{\simeq} -\frac{1}{\Gamma(1-\alpha)t^\alpha}\int_0^tdt'   \dot{x}(t')\simeq \frac{x_0}{\Gamma(1-\alpha)t^\alpha}.\label{conditional mean of non-gaussian process}
\end{align}
This relation is illustrated in Fig.~\ref{Fig_anharmonic}(b) and is consistent with stochastic simulations. As a consequence, the behavior of the correlation function at long times can be written as
\begin{align}
\frac{\langle x(t)x(0)\rangle}{l^2} = \int dx_0 p_s(x_0) \frac{\langle x(t)x_0\rangle}{l^2}=\int dx_0 p_s(x_0) \frac{x_0^2 }{\Gamma(1-\alpha)t^\alpha l^2} =\frac{1 }{\Gamma(1-\alpha)t^\alpha }, 
\end{align}
which enables us to identify the parameter $A$ in the correlation function.  

 Finally, for this strongly non-Gaussian process, we compare the results for the normalized covariance between the first and second passage times to rare events, $T$ and $T_2$ respectively, with Eq.~(\ref{CovT1T2}) of the main text in Fig.~\ref{Fig_anharmonic}(c). For the comparison, the choice of the target thresholds, $L$ is always greater than unity [see the caption of Fig.~\ref{Fig_anharmonic}(c)], where $U(x)$ deviates significantly from its harmonic counterpart as can be seen from  Fig.~\ref{Fig_anharmonic}(a). The excellent agreement between the theoretical and simulation results suggests that correlations between successive passages induced by long-term memory should obey the same kind of laws even in the non-Gaussian case.

\section{Validity of   approximations}
\label{SectionRelaxApprox} 

Our theory relies on the approximations that (i) the statistics of the paths after the first passage, conditional on the value of $T$, is Gaussian, and (ii) that their covariance can be replaced by the covariance of the process conditional to $x(T)=L$ (instead of $x$ being at $L$ at $T$ for the first time). We show here that these approximations become exact, at order $\varepsilon$, at the timescale $T_{\mathrm{RE}}$. We also explain why one can omit the contribution from the intermediate timescale $\tau_d$ in our theory.

\subsection{Relaxing the approximation for the covariance} 
In this Section, we keep the Gaussian approximation (i) but relax the approximation (ii) for the covariance which we calculate at long timescales. To find a general equation for the covariance, we consider Eq.~(\ref{renewal eqn_nPoints}) for $m=2$:
\begin{align}
p(L, t; x_1, t+t_1;x_2, t+t_2)=\int_0^t d\tau \ F(\tau) \ p(L,t; x_1,t+t_1;x_2,t+t_2| T=\tau). \label{renewal 2 points}
\end{align}
Multiplying by $x_1 x_2$ and integrating over $x_1$ and $x_2$ leads to
\begin{align}
p_s(L) &[\text{Cov}(x(t_1+t)x(t_2+t)\vert x(t)=0)+ \mathbb{E}(x(t_1+t) \vert x(t)=L)\mathbb{E}(x(t_2+t) \vert x(t)=L) ]= \nonumber\\
\int_0^t du  &F(t-u) \ p(L,t  \vert T=t-u)\Big[ \text{cov}(x(t+t_1),x(t+t_2)\vert T=t-u;x(t)=L)\nonumber \\+&\mathbb{E}(x(t+t_1)\vert T=t-u;x(t)=L) \mathbb{E}(x(t+t_2)\vert T=t-u;x(t)=L)\Big] , \label{EqCondCov}
\end{align}
where
\begin{align}
\text{cov}(x(t+t_1),x(t+t_2)\vert T=t-u;x(t)=L)=&\text{cov}(x(t+t_1),x(t+t_2)\vert T=t-u)\nonumber\\
&-\frac{\text{cov}(x(t+t_1),x(t)\vert T=t-u)\text{cov}(x(t),x(t+t_2)\vert T=t-u)}{\text{var}(x(t) \vert T=t-u)}. \label{GenFormCondCov}
\end{align}
Now, for $t_1-t_2=\mathcal{O}(T_{\mathrm{RE}})$, we consider the scaling  ansatz
\begin{align}
\text{cov}(x(t+t_1),x(t+t_2)\vert T=t-u)=\frac{A}{T_{\mathrm{RE}}^\alpha}S\left(\frac{t+t_1}{T_{\mathrm{RE}}},\frac{t+t_2}{T_{\mathrm{RE}}};\frac{t-u}{T_{\mathrm{RE}}}\right),
\end{align}
with $S$ a dimensionless function. We also assume that $\text{var}(x(t+t_1) \vert T=t-u)=l^2 + \mathcal{O}(\varepsilon^2).$
Using this ansatz, we find that the terms in the second line of Eq.~(\ref{GenFormCondCov}) are negligible, so that
\begin{align}
\text{cov}(x(t+t_1),x(t+t_2)\vert T=t-u;x(t)=L)\simeq \frac{A}{T_{\mathrm{RE}}^\alpha}S\left(\frac{t+t_1}{T_{\mathrm{RE}}},\frac{t+t_2}{T_{\mathrm{RE}}};\frac{t-u}{T_{\mathrm{RE}}}\right).\label{ScalingCov}
\end{align}
Now, if one considers Eq.~
(\ref{EqCondCov}) at leading order, we realize that the terms containing products of averages are negligible because of order $\varepsilon^2$, and at leading order one can use $F\simeq F_0$. Evaluating as usual the integral as a contribution where $u\sim t^*$ and another one where $u\sim T_{\mathrm{RE}}$, we obtain
\begin{align}
\frac{1}{\vert \bar{t}_1-\bar{t}_2\vert^{\alpha}}= e^{-\overline{t}}\ S( \bar{t}+\bar{t}_1, \bar{t}+\bar{t}_2; \bar{t})
+\int_0^{\bar{t}} d\bar{u} \  &e^{-(\bar{t}-\bar{u})}  \ S( \bar{t}+\bar{t}_1, \bar{t}+\bar{t}_2; \bar{t}).
\end{align}
This equation admits the trivial solution
\begin{align}
S(  \bar{t}_1,  \bar{t}_2; \bar{t})= {1}/{\vert \bar{t}_1-\bar{t}_2\vert^{\alpha}}, \label{SolS}
\end{align}
which is the same result as one would obtain by using our approximation for the covariance. Now, if the above result holds, with the scaling (\ref{ScalingCov}), one can show that the key equations (\ref{CondMeanSmallu}),
 (\ref{CondMeanLargeu}) and (\ref{DefR}) of the theory are valid, even without invoking any \textit{a priori} approximation for the covariance of the process after the first passage.  
 We therefore conclude that our theory to estimate the distribution of first passage time does not rely on any approximation for the covariance of the process at long timescales.  Of course, this argument does not hold for the covariance at times $t^*$ after the first passage, which could differ from the stationary one, but this would only modify the value of $T_{\mathrm{RE}}$ and not the shape of the distribution. 

\subsection{Relaxing the Gaussian approximation }

We now investigate the validity of the Gaussian approximation for the trajectories in the future of the FPT, in the determination of the FPT. Let us  consider  again Eq. (\ref{renewal eqn_nPoints}), which reads
\begin{align}
&p(L,t ; \{x_i, t+t_i\})=\int_0^t du \ F(t-u) \ p(L,t\vert T=t-u) p(\{x_i,t+t_i\}|T=t-u,x(t)=L), 
\end{align}
where we have used the notation $\{x_i,t+t_i\}$ as a shortcut for $x_1,t+t_1;x_2,t+t_2,...,x_n,t+t_n$. The above equation is  exact for any $0<t_1<...<t_n$  and any $x_1,...,x_n$. Since $n$ and the paths $\{x_i,t_i\}$ are abitrary, this equation defines the statistics of trajectories after the FPT and the distribution of FPTs. {Here, we show that, assuming a Gaussian form for the paths after the FPT,  the Eq.~(\ref{renewal eqn_nPoints}) is satisfied at next-to-leading order in $\varepsilon$ for arbitrary integer $n$, when all $t_i=\overline{t}_iT_{\mathrm{RE}}$ are of order $T_{\mathrm{RE}}$, which  suggests that our Gaussian ansatz is exact at  order $\varepsilon$ at this timescale $T_{\mathrm{RE}}$.}

We note that $p(L,t ; \{x_i,t+t_i\})$ does not depend on $t$, and that
\begin{align}
p(L,t ; \{x_i,t+t_i\})=p_s(L)p( \{x_i,t+t_i\} \vert x(t)=L). \label{R0214}
\end{align}
Hence,  combining  (\ref{renewal eqn_nPoints}) and Eq.~(\ref{RenewalEq}), one obtains
\begin{align}
0=
\int_0^t du F(t-u) p(L,t\vert T=t-u) [p(\{x_i,t+t_i\}|T=t-u,x(t)=L)
-p( \{x_i,t+t_i\}\vert x(t)=L)].
\end{align}
Now, in the case that all times $t_i=\overline{t}_iT_{\mathrm{RE}}$ are of order $T_{\mathrm{RE}}$, we analyze this equation as in the end matter, with a contribution from $u\sim t^*$, and a contribution for $u\sim T_{\mathrm{RE}}$:
\begin{align}
&0=
\frac{t^*c_H}{l^*T_{\mathrm{RE}}} F_0(\overline{t})  [p(\{x_i,t+t_i\} |T=t)
-p( \{x_i,t+t_i\} \vert x(t)=L)]\nonumber\\
&+ \int_0^{\overline{t}} d\overline{u}  F_0(\overline{t}-\overline{u})  p(L,t\vert T=t-u)
 [p(\{x_i,t+t_i\}|T=t-u,x(t)=L)
-p(\{x_i,t+t_i\}\vert x(t)=L)].\label{40211}
\end{align}
Note that for the contribution with $u= U t^*$, we have used the fact that the conditions $T=t-u$ and $x(t)=L$ are redundant for small enough $u$. 
We can also write the above equation as
\begin{align}
&0=
p_s(L) F_0(\overline{t})  [p(\{x_i,t+t_i\}|T=t)
-p( \{x_i,t+t_i\} \vert x(t)=L)]\nonumber\\
&+ \int_0^{\overline{t}} d\overline{u} \  F_0(\overline{t}-\overline{u}) [ p(L,t  , \{x_i,t+t_i\}|T=t-u) -p(L,t \vert T=t-u) p( \{x_i,t+t_i\}\vert x(t)=L)].
\end{align}
Now, let us define 
\begin{align}
W_n(x_1,\overline{t}_1;...;x_n,\overline{t}_n;\overline{t})=p(\{x_i,t+t_i\}\vert T=t)-p(\{x_i,t+t_i\}), \label{DefWn}
\end{align}
where $\overline{t}_i=t_i/T_{\mathrm{RE}}$. 
Noting that $p(L,t|T=t-u)=p_s(L)+W_1(L,\overline{u};\overline{t}-\overline{u})$,
and using  Eq.~(\ref{R0214}), Eq.~(\ref{40211}) becomes 
\begin{align}
 &0=p_s(L) F_0(\overline{t})  \left[W_n(x_1,\overline{t}_1;...;x_n,\overline{t}_n;\overline{t})+p( x_1, t+t_1;...;x_n, t+t_n )-p( x_1, t+t_1;...;x_n, t+t_n \vert x(t)=L)\right]
\nonumber\\
& + \int_0^{\overline{t}} d\overline{u}    F_0(\overline{t}-\overline{u}) \Big[ W_{n+1}(L,\overline{u}  ;x_1,\overline{u}+\overline{t}_1;...;x_n,\overline{u}+\overline{t}_n;\overline{t}-\overline{u}) 
 -W_1(L,\overline{u};\overline{t}-\overline{u}) p(x_1,t+t_1; \ldots; x_n,t+t_n\vert x(t)=L)\Big]. \label{renewal eqn in terms of W}
\end{align}
Now, we note that
\begin{align}
p( x_1, t+t_1;...;x_n, t+t_n )=\frac{1}{\sqrt{\det{2\pi \ve[M]_n}}}e^{-\ve[x]^{\mathbb{T}}\cdot \ve[M]_n^{-1}\cdot\ve[x]/2}\equiv p_s(\ve[x]),
\end{align}
where the notation $p_s(\ve[x])$ is defined by in the above equation, $\ve[x]^{\mathbb{T}}=(x_1,...,x_n)$ and $\ve[M]_n$ is a $n\times n$ matrix with elements $(\ve[M]_n)_{ij}=l^2\phi(t_i-t_j)$. 
Using formulas for conditional Gaussian distributions \cite{Eaton1983}, we can write 
\begin{align}
p( x_1, t+t_1;...;x_n, t+t_n \vert x(t)=L)=\frac{1}{\sqrt{\det{2\pi \ve[M]^*_n}}}e^{-(\ve[x]-\ve[m])^{\mathbb{T}}\cdot (\ve[M]^*_n)^{-1}\cdot(\ve[x]-\ve[m])/2}
\end{align}
with $(\ve[M]_n^*)_{ij}=(\ve[M]_n)_{ij}-l^2\phi(t_i)\phi(t_j)$ and $ (\ve[m])_i=L \phi(t_i)$. 
When all $t_i=\mathcal{O}(T_{\mathrm{RE}})$, we have $\ve[M]_n^* \simeq \ve[M]_n$ at order $\varepsilon$, and 
\begin{align}
&p( x_1, t+t_1;...;x_n, t+t_n \vert x(t)=L) -p( x_1, t+t_1;...;x_n, t+t_n )
\simeq  p_s(\ve[x])\ \ve[x]^{\mathbb{T}}\cdot\ve[M]_n^{-1}\cdot\ve[m]. 
\end{align}
  Since $(\ve[M]_n)_{ij}\simeq \delta_{ij}l^2$ at leading order in $\varepsilon$, we have 
\begin{align} 
&p( x_1, t+t_1;...;x_n, t+t_n \vert x(t)=L)-p( x_1, t+t_1;...;x_n, t+t_n ) \simeq  \varepsilon\ p_s(\ve[x])\    \sum^{n}_{i=1} \frac{\tilde{x}_i}{\overline{t}^{\alpha}_i}, \label{EvalDeltap}
\end{align}
where $\tilde{x}_i=x_i/L$.  Also, using the Gaussian ansatz for the process after the FPT, we can write
\begin{align}
p(x_1,t+t_1,...,x_n,t+t_n\vert T=t)=\frac{1}{\sqrt{\det{2\pi \ve[\Omega]_n}}}e^{-\frac{1}{2}\left(\ve[x]-\frac{L A\ve[R]}{T^{\alpha}_{\mathrm{RE}}} \right)^{\mathbb{T}}\cdot (\ve[\Omega]_n)^{-1}\cdot\left(\ve[x]-\frac{L A\ve[R]}{T^{\alpha}_{\mathrm{RE}}} \right)},
\end{align}
where $(\ve[R])_{i}=R(\overline{t}_i,\overline{t})$ [see Eq.~(\ref{muAtScaleTRE})] and, using Eq.~(\ref{SolS}),
\begin{align}
(\ve[\Omega]_n)_{ij}&=\text{cov}(x(t_i),x(t_j)\vert T=t)\simeq \begin{cases}
l^2 \phi(|t_i-t_j|)& (i\ne j),\\
l^2 & (i=j).
\end{cases}
\end{align}
As a result, $\ve[\Omega]_n\simeq \ve[M]_n$ at next-to-leading order in $\varepsilon$, and we can write
\begin{align}
p(x_1,t+t_1,...,x_n,t+t_n\vert T=t) \simeq p_s(\ve[x])\ \left(1+ \varepsilon \sum_{i=1}^n  \tilde{x}_i R(\overline{t}_i,\overline{t}) \right). 
\end{align}
We use this expression to estimate  $W_n$ from Eq. (\ref{DefWn}), leading to 
\begin{align}
W_n(x_1,t_1; \ldots; x_n, t_n;t)& \simeq \varepsilon \ p_s(\ve[x]) \sum_{i=1}^n  \tilde{x}_i R(\overline{t}_i,\overline{t}). \label{Wn}
\end{align}
Applying this expression for one additional variable, we have 
\begin{align}
W_{n+1}(L,{u}  ; x_1,{u}+{t}_1;...;x_n,{u}+{t}_n;{t}-{u})= \varepsilon  p_s(L)p_s(\ve[x]) \left[ R(\overline{u}, \overline{t}-\overline{u}) + \sum_{i=1}^n \tilde{x}_i R(\overline{u}+\overline{t}_i,\overline{t}-\overline{u})  \right], \label{WnPlus1}
\end{align}
where we have used that, at leading order in $\varepsilon$, $p(L, {t}  ; x_1, {t}+ {t}_1;...;x_n, {t}+ {t}_n)\simeq p_s(L)p_s(\ve[x])$. 
Now, using Eqs~(\ref{EvalDeltap}), (\ref{Wn}), (\ref{WnPlus1}), and collecting all terms of order $\varepsilon p_s(L) p_s(\ve[x])$, we obtain
\begin{align}
  e^{-\overline{t}}  \sum_{i=1}^n  \tilde{x}_i \left[R(\overline{t}_i,\overline{t}) -   \frac{1}{\overline{t}^{\alpha}_i}  \right]
 + \int^{\overline{t}}_0 d \overline{u} \ e^{-(\overline{t}-\overline{u})}   \sum_{i=1}^n \tilde{x}_i R(\overline{u}+\overline{t}_i,\overline{t}-\overline{u})    =0.\label{nPointRenEqOrderEpsilon}
\end{align}
This equation is the expression at order $\varepsilon$ of the $n-$point renewal equation (\ref{renewal eqn_nPoints}) when one has used the Gaussian ansatz for the trajectories after the FPT. However, $R$ is already determined from the $1$-point renewal equation. Now, we remark that, $R$ being the solution of (\ref{self consistent eqn for R}), the above equation (\ref{nPointRenEqOrderEpsilon}) is also satisfied, for all $n\ge1$ and all $\tilde{x}_i$. This shows that our Gaussian ansatz for statistics of trajectories after the FPT still enables us to satisfy the $n-$point renewal equation at non-trivial order in $\varepsilon$ at timescales $T_{\mathrm{RE}}$, suggesting that it is exact at this timescale.

 \subsection{Justification for the omission of the timescale 
 $\tau_d$ in the theory}
 \label{JustifOmitTaud}

Here, we justify why, in our analysis, only the timescales $t^*$ and $T_{\mathrm{RE}}$ need to be taken into account, while the behavior of the propagator at scale $\tau_d$ can be neglected. Here, $\tau_d$ represents the characteristic timescale at which the process relax (for example, for the passive process, $\tau_d=(K_\alpha/k)^{1/\alpha}$ and is considered as being of order $1$ since it does not tend to zero or infinity when $L\to\infty$. 

Let us characterize the average trajectories $\mu(t\vert t-u)$ for $u\sim \tau_d$, we use the ansatz 
\begin{align}
\mu(t\vert t-u)\simeq L G(u,t/T_{\text{RE}}). \label{AnsatzTauD}
\end{align}
This is justified because in this case Eqs.~(\ref{muAtScaletStar}) and (\ref{AnsatzTauD}) will predict the same values for $t^*\ll u\ll \tau_d$ at the condition  $G(u,\overline{t})\simeq 1-c u^{2H}$ for ${u}\to0$. We can use the above ansatz to obtain $G$ by combining it with Eqs. (\ref{RenewalEq}), (\ref{EqCond}) and  (\ref{AvTrajStat}) for $t_1=\mathcal{O}(\tau_d)$, noting that the integrals are essentially dominated by small values of $u\sim t^*$, so that
\begin{align}
F(\overline{t}) L[G(t_1,\overline{t})-\phi(t_1)] \frac{t^*}{l^*} c_H  =0, 
\end{align}
which leads to the simple solution $G(t_1,\overline{t})=\phi(t_1) $. 
Note that Eqs. (\ref{AnsatzTauD}) and (\ref{muAtScaleTRE}) will predict the same value the same value of $\mu$ in the regime $\tau_d\ll u \ll T_{\mathrm{RE}}$ at the condition $R(\overline{u},\overline{\tau}) \simeq \overline{u}^{-\alpha}$ for $\overline{u}\to0$. 

We consider the example of obtaining the renewal equation (\ref{RenewalEq}) at order $\varepsilon$. We introduce intermediate timescales $\lambda_1$ and $\lambda_2$ such that 
\begin{align}
t^*\ll \lambda_1\ll \tau_d\ll \lambda_2\ll T_{\mathrm{RE}}, \label{OrderLambda12}
\end{align}
and we write the integrals appearing in the renewal equation (\ref{RenewalEq}) as $\int_0^{\overline{t}T_{\mathrm{RE}}}(...)=\int_0^{\lambda_1}(...)+\int_{\lambda_1}^{\lambda_2}(...)+\int_{\lambda_2}^{\overline{t}T_{\mathrm{RE}}}(...)$. In each interval, we use the appropriate form of the propagator at scale $t^*,\tau_d,T_{\mathrm{RE}}$, respectively deduced from $p\propto q^*$ for $u\sim t^*$, Eq.~(\ref{AnsatzTauD}) with $G(t_1,\overline{t})=\phi(t_1)$ for $t\sim \tau_d$, and Eq. (\ref{DefR}) for $t\sim T_{\mathrm{RE}}$:
\begin{align}
p_s(L)= \overline{F}(\overline{t})\frac{ \int^{\lambda_1/t^*}_0 dU q^*(U) t^*}{l^*T_{\mathrm{RE}}}+ \int_{\lambda_1}^{\lambda_2} \frac{du\overline{F}(\overline{t})e^{-\frac{L^2[1-\phi(u)]^2}{2l^2[1-\phi^2(u)]}}}{T_{\mathrm{RE}}\sqrt{2\pi l^2[1-\phi^2(u)]}}
+\int_{\lambda_2/T_{\mathrm{RE}}}^{\overline{t}} d\overline{u} \overline{F}(\overline{t}-\overline{u}) p_s(L)[1+\varepsilon R(\overline{u},\overline{t}-\overline{u})], \label{04931}
\end{align}
where for brevity we have written  $\overline{F}=F_0+\varepsilon F_1$. 
Now, for $\lambda_2/T_{\mathrm{RE}}\ll 1$, one evaluates
\begin{align}
\int_0^{\lambda_2/T_{\mathrm{RE}}} d\overline{u} \overline{F}(\overline{t}-\overline{u})[1+\varepsilon R(\overline{u},\overline{t}-\overline{u})]\simeq\int_0^{\lambda_2/T_{\mathrm{RE}}} d\overline{u}\overline{F}(\overline{t})  \left(1+\frac{\varepsilon}{\overline{u}^\alpha}\right)\simeq\overline{F}(\overline{t})\left(\frac{\lambda_2}{T_{\mathrm{RE}}}+\frac{\varepsilon\ \lambda_2^{1-\alpha}}{(1-\alpha)T_{\mathrm{RE}}^{1-\alpha}}\right). \nonumber
\end{align}
Next, we evaluate Eq.~(\ref{04931}), where we use $\lambda_1\gg1$ in the first integral of the right hand side, and $\lambda_2/T_{\mathrm{RE}}\ll 1$ in the third integral, where we use the trick $\int_{\lambda_2/T_{\mathrm{RE}}}^{\overline{t}}(...)=\int_{0}^{\overline{t}}(...)-\int_0^{\lambda_2/T_{\mathrm{RE}}}(...)$ and the above expression. We obtain
\begin{align}
p_s(L)= \overline{F}(\overline{t}) \int^{\infty}_0 dU\ q^*(U)\frac{t^*}{l^*T_{\mathrm{RE}}}
+\int_{0}^{\overline{t}} d\overline{u}\ \overline{F}(\overline{t}-\overline{u}) p_s(L)\left(1+\varepsilon R(\overline{u},\overline{t}-\overline{u})\right)+\mathcal{E}, \label{RenEqWithError}
\end{align}
where one recovers Eq.~(\ref{eqn for h and R}), except for an ``error term'' $\mathcal{E}$ that reads
\begin{align}
\mathcal{E} = \overline{F}(\overline{t}) \left(\int_{\lambda_1}^{\lambda_2} \  \frac{du\ e^{-L^2\frac{(1-\phi(u))^2}{2l^2[1-\phi^2(u)]}}}{T_{\mathrm{RE}}\sqrt{2\pi l^2(1-\phi^2(u))}} -p_s(L)\frac{\lambda_2}{T_{\mathrm{RE}}}-p_s(L)\varepsilon\frac{\lambda_2^{1-\alpha}}{(1-\alpha)T_{\mathrm{RE}}^{1-\alpha}}\right). \label{Error}
\end{align}
We evaluate $\mathcal{E}$ for large $L$, taking into account the order of magnitude of $\lambda_1,\lambda_2$ in Eq.~(\ref{OrderLambda12}). The main contribution to the integral in (\ref{Error})  comes from the values of $u$ close to $\lambda_1$, which give contributions proportional to $e^{-(\lambda_1/t^*)^{2H}}/T_{\mathrm{RE}}$, which are exponentially small compared to $p_s(L)$ and therefore do not contribute in Eq.~(\ref{RenEqWithError}). Next, the integral in (\ref{Error}) diverges for large $\lambda_2$, evaluating the divergences leads to $\mathcal{E}\sim  p_s(L) \varepsilon (\lambda_2/T_{\mathrm{RE}})^{1-\alpha}$, which is small compared to terms of the order $p_s(L)\varepsilon$ when $\lambda_2\ll T_{\mathrm{RE}}$. We conclude that it is safe to neglect $\mathcal{E}$ in (\ref{RenEqWithError}), so that the contributions due to the timescales $\tau_d$ in the renewal equation are not relevant to investigate the effect of long-term memory.

\bibliographystyle{naturemag}

\begin{thebibliography}{10}
\expandafter\ifx\csname url\endcsname\relax
  \def\url#1{\texttt{#1}}\fi
\expandafter\ifx\csname urlprefix\endcsname\relax\def\urlprefix{URL }\fi
\providecommand{\bibinfo}[2]{#2}
\providecommand{\eprint}[2][]{\url{#2}}

\bibitem{hanggi1990reaction}
\bibinfo{author}{H{\"a}nggi, P.}, \bibinfo{author}{Talkner, P.} \&
  \bibinfo{author}{Borkovec, M.}
\newblock \bibinfo{title}{Reaction-rate theory: fifty years after kramers}.
\newblock \emph{\bibinfo{journal}{Rev. Mod. Phys.}}
  \textbf{\bibinfo{volume}{62}}, \bibinfo{pages}{251} (\bibinfo{year}{1990}).

\bibitem{pollak2005reaction}
\bibinfo{author}{Pollak, E.} \& \bibinfo{author}{Talkner, P.}
\newblock \bibinfo{title}{Reaction rate theory: What it was, where is it today,
  and where is it going?}
\newblock \emph{\bibinfo{journal}{Chaos: An Interdisciplinary Journal of
  Nonlinear Science}} \textbf{\bibinfo{volume}{15}}, \bibinfo{pages}{026116}
  (\bibinfo{year}{2005}).

\bibitem{chowdhury2022extreme}
\bibinfo{author}{Chowdhury, S.~N.}, \bibinfo{author}{Ray, A.},
  \bibinfo{author}{Dana, S.~K.} \& \bibinfo{author}{Ghosh, D.}
\newblock \bibinfo{title}{Extreme events in dynamical systems and random
  walkers: A review}.
\newblock \emph{\bibinfo{journal}{Physics Reports}}
  \textbf{\bibinfo{volume}{966}}, \bibinfo{pages}{1--52}
  (\bibinfo{year}{2022}).

\bibitem{bullerjahn2014theory}
\bibinfo{author}{Bullerjahn, J.~T.}, \bibinfo{author}{Sturm, S.} \&
  \bibinfo{author}{Kroy, K.}
\newblock \bibinfo{title}{Theory of rapid force spectroscopy}.
\newblock \emph{\bibinfo{journal}{Nat. Comm.}} \textbf{\bibinfo{volume}{5}}
  (\bibinfo{year}{2014}).

\bibitem{bullerjahn2020non}
\bibinfo{author}{Bullerjahn, J.~T.}, \bibinfo{author}{Sturm, S.} \&
  \bibinfo{author}{Kroy, K.}
\newblock \bibinfo{title}{Non-markov bond model for dynamic force
  spectroscopy}.
\newblock \emph{\bibinfo{journal}{J. Chem. Phys.}}
  \textbf{\bibinfo{volume}{152}}, \bibinfo{pages}{064104}
  (\bibinfo{year}{2020}).

\bibitem{bullerjahn2016analytical}
\bibinfo{author}{Bullerjahn, J.} \& \bibinfo{author}{Kroy, K.}
\newblock \bibinfo{title}{Analytical catch-slip bond model for arbitrary forces
  and loading rates}.
\newblock \emph{\bibinfo{journal}{Phys. Rev. E}} \textbf{\bibinfo{volume}{93}},
  \bibinfo{pages}{012404} (\bibinfo{year}{2016}).

\bibitem{jeppesen2001impact}
\bibinfo{author}{Jeppesen, C.} \emph{et~al.}
\newblock \bibinfo{title}{Impact of polymer tether length on multiple
  ligand-receptor bond formation}.
\newblock \emph{\bibinfo{journal}{Science}} \textbf{\bibinfo{volume}{293}},
  \bibinfo{pages}{465--468} (\bibinfo{year}{2001}).

\bibitem{Badoual2002a}
\bibinfo{author}{Badoual, M.}, \bibinfo{author}{J\"ulicher, F.} \&
  \bibinfo{author}{Prost, J.}
\newblock \bibinfo{title}{Bidirectional cooperative motion of molecular
  motors.}
\newblock \emph{\bibinfo{journal}{Proc Natl Acad Sci USA}}
  \textbf{\bibinfo{volume}{99}}, \bibinfo{pages}{6696--6701}
  (\bibinfo{year}{2002}).

\bibitem{guerin2011}
\bibinfo{author}{Gu{\'e}rin, T.}, \bibinfo{author}{Prost, J.} \&
  \bibinfo{author}{Joanny, J.-F.}
\newblock \bibinfo{title}{Motion reversal of molecular motor assemblies due to
  weak noise}.
\newblock \emph{\bibinfo{journal}{Phys Rev Lett}}
  \textbf{\bibinfo{volume}{106}}, \bibinfo{pages}{068101}
  (\bibinfo{year}{2011}).

\bibitem{Bouchaud1998}
\bibinfo{author}{Bouchaud, J.~P.} \& \bibinfo{author}{Cont, R.}
\newblock \bibinfo{title}{A Langevin approach to stock market fluctuations and
  crashes}.
\newblock \emph{\bibinfo{journal}{Eur Phys J B}} \textbf{\bibinfo{volume}{6}},
  \bibinfo{pages}{543--550} (\bibinfo{year}{1998}).

\bibitem{ragone2018computation}
\bibinfo{author}{Ragone, F.}, \bibinfo{author}{Wouters, J.} \&
  \bibinfo{author}{Bouchet, F.}
\newblock \bibinfo{title}{Computation of extreme heat waves in climate models
  using a large deviation algorithm}.
\newblock \emph{\bibinfo{journal}{Proc. Natl. Acad. Sci. USA}}
  \textbf{\bibinfo{volume}{115}}, \bibinfo{pages}{24--29}
  (\bibinfo{year}{2018}).

\bibitem{kamenev2008colored}
\bibinfo{author}{Kamenev, A.}, \bibinfo{author}{Meerson, B.} \&
  \bibinfo{author}{Shklovskii, B.}
\newblock \bibinfo{title}{How colored environmental noise affects population
  extinction}.
\newblock \emph{\bibinfo{journal}{Phys. Rev. Lett.}}
  \textbf{\bibinfo{volume}{101}}, \bibinfo{pages}{268103}
  (\bibinfo{year}{2008}).

\bibitem{dykman2008disease}
\bibinfo{author}{Dykman, M.~I.}, \bibinfo{author}{Schwartz, I.~B.} \&
  \bibinfo{author}{Landsman, A.~S.}
\newblock \bibinfo{title}{Disease extinction in the presence of random
  vaccination}.
\newblock \emph{\bibinfo{journal}{Phys. Rev. Lett.}}
  \textbf{\bibinfo{volume}{101}}, \bibinfo{pages}{078101}
  (\bibinfo{year}{2008}).

\bibitem{Freidlin1984}
\bibinfo{author}{Freidlin, M.~I.} \& \bibinfo{author}{Wentzell, A.~D.}
\newblock \emph{\bibinfo{title}{Random Perturbations of Dynamical Systems}}
  (\bibinfo{publisher}{Springer-Verlage, New-York, Berlin},
  \bibinfo{year}{1984}).

\bibitem{Maier1992}
\bibinfo{author}{Maier, R.} \& \bibinfo{author}{Stein, D.}
\newblock \bibinfo{title}{Transition-rate theory for nongradient drift fields.}
\newblock \emph{\bibinfo{journal}{Phys Rev Lett}}
  \textbf{\bibinfo{volume}{69}}, \bibinfo{pages}{3691--3695}
  (\bibinfo{year}{1992}).

\bibitem{bouchet2016generalisation}
\bibinfo{author}{Bouchet, F.} \& \bibinfo{author}{Reygner, J.}
\newblock \bibinfo{title}{Generalisation of the eyring--kramers transition rate
  formula to irreversible diffusion processes}.
\newblock In \emph{\bibinfo{booktitle}{Annales Henri Poincar{\'e}}},
  vol.~\bibinfo{volume}{17}, \bibinfo{pages}{3499--3532}
  (\bibinfo{organization}{Springer}, \bibinfo{year}{2016}).

\bibitem{delacruz2018minimum}
\bibinfo{author}{de~la Cruz, R.}, \bibinfo{author}{Perez-Carrasco, R.},
  \bibinfo{author}{Guerrero, P.}, \bibinfo{author}{Alarcon, T.} \&
  \bibinfo{author}{Page, K.~M.}
\newblock \bibinfo{title}{Minimum action path theory reveals the details of
  stochastic transitions out of oscillatory states}.
\newblock \emph{\bibinfo{journal}{Phys. Rev. Lett.}}
  \textbf{\bibinfo{volume}{120}}, \bibinfo{pages}{128102}
  (\bibinfo{year}{2018}).

\bibitem{santhanam2008return}
\bibinfo{author}{Santhanam, M.} \& \bibinfo{author}{Kantz, H.}
\newblock \bibinfo{title}{Return interval distribution of extreme events and
  long-term memory}.
\newblock \emph{\bibinfo{journal}{Phys. Rev. E}} \textbf{\bibinfo{volume}{78}},
  \bibinfo{pages}{051113} (\bibinfo{year}{2008}).

\bibitem{bunde2005long}
\bibinfo{author}{Bunde, A.}, \bibinfo{author}{Eichner, J.~F.},
  \bibinfo{author}{Kantelhardt, J.~W.} \& \bibinfo{author}{Havlin, S.}
\newblock \bibinfo{title}{Long-term memory: A natural mechanism for the
  clustering of extreme events and anomalous residual times in climate
  records}.
\newblock \emph{\bibinfo{journal}{Phys. Rev. Lett.}}
  \textbf{\bibinfo{volume}{94}}, \bibinfo{pages}{048701}
  (\bibinfo{year}{2005}).

\bibitem{Panja2010}
\bibinfo{author}{Panja, D.}
\newblock \bibinfo{title}{Anomalous polymer dynamics is non-markovian: memory
  effects and the generalized langevin equation formulation}.
\newblock \emph{\bibinfo{journal}{J. Stat. Mech.: Theor. Exp.}}
  \textbf{\bibinfo{volume}{2010}}, \bibinfo{pages}{P06011}
  (\bibinfo{year}{2010}).

\bibitem{kou2004generalized}
\bibinfo{author}{Kou, S.} \& \bibinfo{author}{Xie, X.~S.}
\newblock \bibinfo{title}{Generalized langevin equation with fractional
  gaussian noise: subdiffusion within a single protein molecule}.
\newblock \emph{\bibinfo{journal}{Phys. Rev. Lett.}}
  \textbf{\bibinfo{volume}{93}}, \bibinfo{pages}{180603}
  (\bibinfo{year}{2004}).

\bibitem{Min2005}
\bibinfo{author}{Min, W.}, \bibinfo{author}{Luo, G.},
  \bibinfo{author}{Cherayil, B.~J.}, \bibinfo{author}{Kou, S.} \&
  \bibinfo{author}{Xie, X.~S.}
\newblock \bibinfo{title}{Observation of a power-law memory kernel for
  fluctuations within a single protein molecule}.
\newblock \emph{\bibinfo{journal}{Phys. Rev. Lett.}}
  \textbf{\bibinfo{volume}{94}}, \bibinfo{pages}{198302}
  (\bibinfo{year}{2005}).

\bibitem{granek2005fractons}
\bibinfo{author}{Granek, R.} \& \bibinfo{author}{Klafter, J.}
\newblock \bibinfo{title}{Fractons in proteins: Can they lead to anomalously
  decaying time autocorrelations?}
\newblock \emph{\bibinfo{journal}{Physical review letters}}
  \textbf{\bibinfo{volume}{95}}, \bibinfo{pages}{098106}
  (\bibinfo{year}{2005}).

\bibitem{yang2003protein}
\bibinfo{author}{Yang, H.} \emph{et~al.}
\newblock \bibinfo{title}{Protein conformational dynamics probed by
  single-molecule electron transfer}.
\newblock \emph{\bibinfo{journal}{Science}} \textbf{\bibinfo{volume}{302}},
  \bibinfo{pages}{262--266} (\bibinfo{year}{2003}).

\bibitem{ReviewBray}
\bibinfo{author}{Bray, A.~J.}, \bibinfo{author}{Majumdar, S.~N.} \&
  \bibinfo{author}{Schehr, G.}
\newblock \bibinfo{title}{Persistence and first-passage properties in
  nonequilibrium systems}.
\newblock \emph{\bibinfo{journal}{Adv. Phys.}} \textbf{\bibinfo{volume}{62}},
  \bibinfo{pages}{225--361} (\bibinfo{year}{2013}).

\bibitem{zhong2018generalized}
\bibinfo{author}{Zhong, W.}, \bibinfo{author}{Panja, D.},
  \bibinfo{author}{Barkema, G.~T.} \& \bibinfo{author}{Ball, R.~C.}
\newblock \bibinfo{title}{Generalized langevin equation formulation for
  anomalous diffusion in the ising model at the critical temperature}.
\newblock \emph{\bibinfo{journal}{Phys. Rev. E}} \textbf{\bibinfo{volume}{98}},
  \bibinfo{pages}{012124} (\bibinfo{year}{2018}).

\bibitem{sarkar2025tracer}
\bibinfo{author}{Sarkar, R.} \& \bibinfo{author}{Santra, I.}
\newblock \bibinfo{title}{Tracer dynamics in an interacting active bath:
  fluctuations and energy partition}.
\newblock \emph{\bibinfo{journal}{New Journal of Physics}}
  \textbf{\bibinfo{volume}{27}}, \bibinfo{pages}{094601}
  (\bibinfo{year}{2025}).

\bibitem{middleton2003firing}
\bibinfo{author}{Middleton, J.}, \bibinfo{author}{Chacron, M.~J.},
  \bibinfo{author}{Lindner, B.} \& \bibinfo{author}{Longtin, A.}
\newblock \bibinfo{title}{Firing statistics of a neuron model driven by
  long-range correlated noise}.
\newblock \emph{\bibinfo{journal}{Physical Review E}}
  \textbf{\bibinfo{volume}{68}}, \bibinfo{pages}{021920}
  (\bibinfo{year}{2003}).

\bibitem{lennartz2008long}
\bibinfo{author}{Lennartz, S.}, \bibinfo{author}{Livina, V.},
  \bibinfo{author}{Bunde, A.} \& \bibinfo{author}{Havlin, S.}
\newblock \bibinfo{title}{Long-term memory in earthquakes and the distribution
  of interoccurrence times}.
\newblock \emph{\bibinfo{journal}{Europhys. Lett.}}
  \textbf{\bibinfo{volume}{81}}, \bibinfo{pages}{69001} (\bibinfo{year}{2008}).

\bibitem{bunde2013there}
\bibinfo{author}{Bunde, A.}, \bibinfo{author}{B{\"u}ntgen, U.},
  \bibinfo{author}{Ludescher, J.}, \bibinfo{author}{Luterbacher, J.} \&
  \bibinfo{author}{Von~Storch, H.}
\newblock \bibinfo{title}{Is there memory in precipitation?}
\newblock \emph{\bibinfo{journal}{Nature Climate Change}}
  \textbf{\bibinfo{volume}{3}}, \bibinfo{pages}{174--175}
  (\bibinfo{year}{2013}).

\bibitem{pickands1969upcrossing}
\bibinfo{author}{Pickands, J.}
\newblock \bibinfo{title}{Upcrossing probabilities for stationary gaussian
  processes}.
\newblock \emph{\bibinfo{journal}{Transactions of the American Mathematical
  Society}} \textbf{\bibinfo{volume}{145}}, \bibinfo{pages}{51--73}
  (\bibinfo{year}{1969}).

\bibitem{pickands1969asymptotic}
\bibinfo{author}{Pickands, J.}
\newblock \bibinfo{title}{Asymptotic properties of the maximum in a stationary
  gaussian process.}
\newblock \emph{\bibinfo{journal}{Transactions of the American Mathematical
  Society}} \textbf{\bibinfo{volume}{145}}, \bibinfo{pages}{75--86}
  (\bibinfo{year}{1969}).

\bibitem{carpenter2022long}
\bibinfo{author}{Carpenter, S.~R.}, \bibinfo{author}{Gahler, M.~R.},
  \bibinfo{author}{Kucharik, C.~J.} \& \bibinfo{author}{Stanley, E.~H.}
\newblock \bibinfo{title}{Long-range dependence and extreme values of
  precipitation, phosphorus load, and cyanobacteria}.
\newblock \emph{\bibinfo{journal}{Proceedings of the National Academy of
  Sciences}} \textbf{\bibinfo{volume}{119}}, \bibinfo{pages}{e2214343119}
  (\bibinfo{year}{2022}).

\bibitem{min2006kramers}
\bibinfo{author}{Min, W.} \& \bibinfo{author}{Xie, X.~S.}
\newblock \bibinfo{title}{Kramers model with a power-law friction kernel:
  Dispersed kinetics and dynamic disorder of biochemical reactions}.
\newblock \emph{\bibinfo{journal}{Phys. Rev. E}} \textbf{\bibinfo{volume}{73}},
  \bibinfo{pages}{010902} (\bibinfo{year}{2006}).

\bibitem{goychuk2009viscoelastic}
\bibinfo{author}{Goychuk, I.}
\newblock \bibinfo{title}{Viscoelastic subdiffusion: From anomalous to normal}.
\newblock \emph{\bibinfo{journal}{Phys. Rev. E}} \textbf{\bibinfo{volume}{80}},
  \bibinfo{pages}{046125} (\bibinfo{year}{2009}).

\bibitem{karplus2000aspects}
\bibinfo{author}{Karplus, M.}
\newblock \bibinfo{title}{Aspects of protein reaction dynamics: Deviations from
  simple behavior} (\bibinfo{year}{2000}).

\bibitem{sunney2002single}
\bibinfo{author}{Sunney~Xie, X.}
\newblock \bibinfo{title}{Single-molecule approach to dispersed kinetics and
  dynamic disorder: Probing conformational fluctuation and enzymatic dynamics}.
\newblock \emph{\bibinfo{journal}{The Journal of chemical physics}}
  \textbf{\bibinfo{volume}{117}}, \bibinfo{pages}{11024--11032}
  (\bibinfo{year}{2002}).

\bibitem{lu1998single}
\bibinfo{author}{Lu, H.~P.}, \bibinfo{author}{Xun, L.} \& \bibinfo{author}{Xie,
  X.~S.}
\newblock \bibinfo{title}{Single-molecule enzymatic dynamics}.
\newblock \emph{\bibinfo{journal}{Science}} \textbf{\bibinfo{volume}{282}},
  \bibinfo{pages}{1877--1882} (\bibinfo{year}{1998}).

\bibitem{flomenbom2005stretched}
\bibinfo{author}{Flomenbom, O.} \emph{et~al.}
\newblock \bibinfo{title}{Stretched exponential decay and correlations in the
  catalytic activity of fluctuating single lipase molecules}.
\newblock \emph{\bibinfo{journal}{Proceedings of the National Academy of
  Sciences}} \textbf{\bibinfo{volume}{102}}, \bibinfo{pages}{2368--2372}
  (\bibinfo{year}{2005}).

\bibitem{english2006ever}
\bibinfo{author}{English, B.~P.} \emph{et~al.}
\newblock \bibinfo{title}{Ever-fluctuating single enzyme molecules:
  Michaelis-menten equation revisited}.
\newblock \emph{\bibinfo{journal}{Nature chemical biology}}
  \textbf{\bibinfo{volume}{2}}, \bibinfo{pages}{87--94} (\bibinfo{year}{2006}).

\bibitem{ferrer2021fluid}
\bibinfo{author}{Ferrer, B.~R.}, \bibinfo{author}{Gomez-Solano, J.~R.} \&
  \bibinfo{author}{Arzola, A.~V.}
\newblock \bibinfo{title}{Fluid viscoelasticity triggers fast transitions of a
  brownian particle in a double well optical potential}.
\newblock \emph{\bibinfo{journal}{Phys. Rev. Lett.}}
  \textbf{\bibinfo{volume}{126}}, \bibinfo{pages}{108001}
  (\bibinfo{year}{2021}).

\bibitem{ginot2022barrier}
\bibinfo{author}{Ginot, F.}, \bibinfo{author}{Caspers, J.},
  \bibinfo{author}{Kr{\"u}ger, M.} \& \bibinfo{author}{Bechinger, C.}
\newblock \bibinfo{title}{Barrier crossing in a viscoelastic bath}.
\newblock \emph{\bibinfo{journal}{Phys. Rev. Lett.}}
  \textbf{\bibinfo{volume}{128}}, \bibinfo{pages}{028001}
  (\bibinfo{year}{2022}).

\bibitem{lavacchi2020barrier}
\bibinfo{author}{Lavacchi, L.}, \bibinfo{author}{Kappler, J.} \&
  \bibinfo{author}{Netz, R.~R.}
\newblock \bibinfo{title}{Barrier crossing in the presence of multi-exponential
  memory functions with unequal friction amplitudes and memory times}.
\newblock \emph{\bibinfo{journal}{Europhys. Lett.}}
  \textbf{\bibinfo{volume}{131}}, \bibinfo{pages}{40004}
  (\bibinfo{year}{2020}).

\bibitem{lavacchi2022non}
\bibinfo{author}{Lavacchi, L.}, \bibinfo{author}{Daldrop, J.~O.} \&
  \bibinfo{author}{Netz, R.~R.}
\newblock \bibinfo{title}{Non-arrhenius barrier crossing dynamics of
  non-equilibrium non-markovian systems}.
\newblock \emph{\bibinfo{journal}{Europhys. Lett.}}
  \textbf{\bibinfo{volume}{139}}, \bibinfo{pages}{51001}
  (\bibinfo{year}{2022}).

\bibitem{kappler2018memory}
\bibinfo{author}{Kappler, J.}, \bibinfo{author}{Daldrop, J.~O.},
  \bibinfo{author}{Br{\"u}nig, F.~N.}, \bibinfo{author}{Boehle, M.~D.} \&
  \bibinfo{author}{Netz, R.~R.}
\newblock \bibinfo{title}{Memory-induced acceleration and slowdown of barrier
  crossing}.
\newblock \emph{\bibinfo{journal}{J. Chem. Phys.}}
  \textbf{\bibinfo{volume}{148}}, \bibinfo{pages}{014903}
  (\bibinfo{year}{2018}).

\bibitem{Caraglio2018}
\bibinfo{author}{Caraglio, M.}, \bibinfo{author}{Put, S.},
  \bibinfo{author}{Carlon, E.} \& \bibinfo{author}{Vanderzande, C.}
\newblock \bibinfo{title}{The influence of absorbing boundary conditions on the
  transition path time statistics}.
\newblock \emph{\bibinfo{journal}{Phys. Chem. Chem. Phys.}}
  \textbf{\bibinfo{volume}{20}}, \bibinfo{pages}{25676--25682}
  (\bibinfo{year}{2018}).

\bibitem{carlon2018effect}
\bibinfo{author}{Carlon, E.}, \bibinfo{author}{Orland, H.},
  \bibinfo{author}{Sakaue, T.} \& \bibinfo{author}{Vanderzande, C.}
\newblock \bibinfo{title}{Effect of memory and active forces on transition path
  time distributions}.
\newblock \emph{\bibinfo{journal}{J. Phys. Chem. B}}  (\bibinfo{year}{2018}).

\bibitem{medina2018transition}
\bibinfo{author}{Medina, E.}, \bibinfo{author}{Satija, R.} \&
  \bibinfo{author}{Makarov, D.~E.}
\newblock \bibinfo{title}{Transition path times in non-markovian activated rate
  processes}.
\newblock \emph{\bibinfo{journal}{J. Phys. Chem. B}}  (\bibinfo{year}{2018}).

\bibitem{goychuk2007anomalous}
\bibinfo{author}{Goychuk, I.} \& \bibinfo{author}{H{\"a}nggi, P.}
\newblock \bibinfo{title}{Anomalous escape governed by thermal 1/f noise}.
\newblock \emph{\bibinfo{journal}{Phys. Rev. Lett.}}
  \textbf{\bibinfo{volume}{99}}, \bibinfo{pages}{200601}
  (\bibinfo{year}{2007}).

\bibitem{Sliusarenko2010}
\bibinfo{author}{Sliusarenko, O.~Y.}, \bibinfo{author}{Gonchar, V.~Y.},
  \bibinfo{author}{Chechkin, A.~V.}, \bibinfo{author}{Sokolov, I.~M.} \&
  \bibinfo{author}{Metzler, R.}
\newblock \bibinfo{title}{Kramers-like escape driven by fractional gaussian
  noise}.
\newblock \emph{\bibinfo{journal}{Phys. Rev. E}} \textbf{\bibinfo{volume}{81}},
  \bibinfo{pages}{041119} (\bibinfo{year}{2010}).

\bibitem{arutkin2020extreme}
\bibinfo{author}{Arutkin, M.}, \bibinfo{author}{Walter, B.} \&
  \bibinfo{author}{Wiese, K.~J.}
\newblock \bibinfo{title}{Extreme events for fractional brownian motion with
  drift: Theory and numerical validation}.
\newblock \emph{\bibinfo{journal}{Phys. Rev. E}}
  \textbf{\bibinfo{volume}{102}}, \bibinfo{pages}{022102}
  (\bibinfo{year}{2020}).

\bibitem{levernier2020kinetics}
\bibinfo{author}{Levernier, N.}, \bibinfo{author}{B{\'e}nichou, O.},
  \bibinfo{author}{Voituriez, R.} \& \bibinfo{author}{Gu{\'e}rin, T.}
\newblock \bibinfo{title}{Kinetics of rare events for non-markovian stationary
  processes and application to polymer dynamics}.
\newblock \emph{\bibinfo{journal}{Phys. Rev. Res.}}
  \textbf{\bibinfo{volume}{2}}, \bibinfo{pages}{012057} (\bibinfo{year}{2020}).

\bibitem{delorme2017pickands}
\bibinfo{author}{Delorme, M.}, \bibinfo{author}{Rosso, A.} \&
  \bibinfo{author}{Wiese, K.~J.}
\newblock \bibinfo{title}{Pickands' constant at first order in an expansion
  around brownian motion}.
\newblock \emph{\bibinfo{journal}{J. Phys. A: Math. Theor.}}
  \textbf{\bibinfo{volume}{50}}, \bibinfo{pages}{16LT04}
  (\bibinfo{year}{2017}).

\bibitem{goswami2023effects}
\bibinfo{author}{Goswami, K.} \& \bibinfo{author}{Metzler, R.}
\newblock \bibinfo{title}{Effects of active noise on transition-path dynamics}.
\newblock \emph{\bibinfo{journal}{Journal of Physics: Complexity}}
  \textbf{\bibinfo{volume}{4}}, \bibinfo{pages}{025005} (\bibinfo{year}{2023}).

\bibitem{zanovello2021target}
\bibinfo{author}{Zanovello, L.}, \bibinfo{author}{Caraglio, M.},
  \bibinfo{author}{Franosch, T.} \& \bibinfo{author}{Faccioli, P.}
\newblock \bibinfo{title}{Target search of active agents crossing high energy
  barriers}.
\newblock \emph{\bibinfo{journal}{Physical Review Letters}}
  \textbf{\bibinfo{volume}{126}}, \bibinfo{pages}{018001}
  (\bibinfo{year}{2021}).

\bibitem{ayaz2021non}
\bibinfo{author}{Ayaz, C.} \emph{et~al.}
\newblock \bibinfo{title}{Non-markovian modeling of protein folding}.
\newblock \emph{\bibinfo{journal}{Proc. Natl. Acad. Sc. USA}}
  \textbf{\bibinfo{volume}{118}}, \bibinfo{pages}{e2023856118}
  (\bibinfo{year}{2021}).

\bibitem{singh2019comment}
\bibinfo{author}{Singh, R.}
\newblock \bibinfo{title}{Comment on “anomalous escape governed by thermal
  1/f noise”}.
\newblock \emph{\bibinfo{journal}{Phys. Rev. Lett.}}
  \textbf{\bibinfo{volume}{123}}, \bibinfo{pages}{238901}
  (\bibinfo{year}{2019}).

\bibitem{bullerjahn2017unified}
\bibinfo{author}{Bullerjahn, J.~T.}
\newblock \emph{\bibinfo{title}{A Unified Theory for Single-molecule Force
  Spectroscopy Experiments and Simulations}}.
\newblock Ph.D. thesis, \bibinfo{school}{Universit{\"a}t Leipzig}
  (\bibinfo{year}{2017}).

\bibitem{zwanzig1990rate}
\bibinfo{author}{Zwanzig, R.}
\newblock \bibinfo{title}{Rate processes with dynamical disorder}.
\newblock \emph{\bibinfo{journal}{Accounts of Chemical Research}}
  \textbf{\bibinfo{volume}{23}}, \bibinfo{pages}{148--152}
  (\bibinfo{year}{1990}).

\bibitem{Barbier2024}
\bibinfo{author}{Barbier-Chebbah, A.}, \bibinfo{author}{B\'enichou, O.},
  \bibinfo{author}{Voituriez, R.} \& \bibinfo{author}{Gu\'erin, T.}
\newblock \bibinfo{title}{Long-term memory induced correction to arrhenius
  law}.
\newblock \emph{\bibinfo{journal}{Nat. Com.}} \textbf{\bibinfo{volume}{15}},
  \bibinfo{pages}{7408} (\bibinfo{year}{2024}).

\bibitem{mason1995optical}
\bibinfo{author}{Mason, T.~G.} \& \bibinfo{author}{Weitz, D.}
\newblock \bibinfo{title}{Optical measurements of frequency-dependent linear
  viscoelastic moduli of complex fluids}.
\newblock \emph{\bibinfo{journal}{Phys. Rev. Lett.}}
  \textbf{\bibinfo{volume}{74}}, \bibinfo{pages}{1250} (\bibinfo{year}{1995}).

\bibitem{gisler1998tracer}
\bibinfo{author}{Gisler, T.} \& \bibinfo{author}{Weitz, D.~A.}
\newblock \bibinfo{title}{Tracer microrheology in complex fluids}.
\newblock \emph{\bibinfo{journal}{Current opinion in colloid \& interface
  science}} \textbf{\bibinfo{volume}{3}}, \bibinfo{pages}{586--592}
  (\bibinfo{year}{1998}).

\bibitem{mason1997particle}
\bibinfo{author}{Mason, T.}, \bibinfo{author}{Ganesan, K.},
  \bibinfo{author}{Van~Zanten, J.}, \bibinfo{author}{Wirtz, D.} \&
  \bibinfo{author}{Kuo, S.}
\newblock \bibinfo{title}{Particle tracking microrheology of complex fluids}.
\newblock \emph{\bibinfo{journal}{Phys. Rev. Lett.}}
  \textbf{\bibinfo{volume}{79}}, \bibinfo{pages}{3282} (\bibinfo{year}{1997}).

\bibitem{Panja2010a}
\bibinfo{author}{Panja, D.}
\newblock \bibinfo{title}{Generalized langevin equation formulation for
  anomalous polymer dynamics}.
\newblock \emph{\bibinfo{journal}{J. Stat. Mech.: Theor. Exp.}}
  \textbf{\bibinfo{volume}{2010}}, \bibinfo{pages}{L02001}
  (\bibinfo{year}{2010}).

\bibitem{bullerjahn2011monomer}
\bibinfo{author}{Bullerjahn, J.~T.}, \bibinfo{author}{Sturm, S.},
  \bibinfo{author}{Wolff, L.} \& \bibinfo{author}{Kroy, K.}
\newblock \bibinfo{title}{Monomer dynamics of a wormlike chain}.
\newblock \emph{\bibinfo{journal}{Europhys. Lett.}}
  \textbf{\bibinfo{volume}{96}}, \bibinfo{pages}{48005} (\bibinfo{year}{2011}).

\bibitem{guerin2016mean}
\bibinfo{author}{Gu{\'e}rin, T.}, \bibinfo{author}{Levernier, N.},
  \bibinfo{author}{B{\'e}nichou, O.} \& \bibinfo{author}{Voituriez, R.}
\newblock \bibinfo{title}{Mean first-passage times of non-markovian random
  walkers in confinement}.
\newblock \emph{\bibinfo{journal}{Nature}} \textbf{\bibinfo{volume}{534}},
  \bibinfo{pages}{356--359} (\bibinfo{year}{2016}).

\bibitem{citeSM}
See Supplemental Material at [URL], which contains   calculation details supporting the results shown in the main text, and a discussion on the validity of the theory and its possible generalization to non-Gaussian processes.   

\bibitem{davies1987tests}
\bibinfo{author}{Davies, R.~B.} \& \bibinfo{author}{Harte, D.}
\newblock \bibinfo{title}{Tests for hurst effect}.
\newblock \emph{\bibinfo{journal}{Biometrika}} \textbf{\bibinfo{volume}{74}},
  \bibinfo{pages}{95--101} (\bibinfo{year}{1987}).

\bibitem{dietrich_fast_1997}
\bibinfo{author}{Dietrich, C.~R.} \& \bibinfo{author}{Newsam, G.~N.}
\newblock \bibinfo{title}{Fast and {Exact} {Simulation} of {Stationary}
  {Gaussian} {Processes} through {Circulant} {Embedding} of the {Covariance}
  {Matrix}}.
\newblock \emph{\bibinfo{journal}{SIAM J. Sci. Comp.}}
  \textbf{\bibinfo{volume}{18}}, \bibinfo{pages}{1088--1107}
  (\bibinfo{year}{1997}).

\bibitem{azencott2018large}
\bibinfo{author}{Azencott, R.}, \bibinfo{author}{Geiger, B.} \&
  \bibinfo{author}{Ott, W.}
\newblock \bibinfo{title}{Large deviations for gaussian diffusions with delay}.
\newblock \emph{\bibinfo{journal}{Journal of Statistical Physics}}
  \textbf{\bibinfo{volume}{170}}, \bibinfo{pages}{254--285}
  (\bibinfo{year}{2018}).

\bibitem{coghi2025accelerated}
\bibinfo{author}{Coghi, F.}, \bibinfo{author}{Duvezin, R.} \&
  \bibinfo{author}{Wettlaufer, J.~S.}
\newblock \bibinfo{title}{Accelerated first-passage dynamics in a non-markovian
  feedback ornstein--uhlenbeck process}.
\newblock \emph{\bibinfo{journal}{Journal of statistical physics}}
  \textbf{\bibinfo{volume}{192}}, \bibinfo{pages}{128} (\bibinfo{year}{2025}).

\bibitem{DataLongMem}%
https://doi.org/10.5281/zenodo.19251535


\bibitem{Eaton1983}
\bibinfo{author}{Eaton, M.~L.}
\newblock \emph{\bibinfo{title}{Multivariate Statistics, A Vector Space
  Approach}}, vol.~\bibinfo{volume}{53} (\bibinfo{publisher}{Institute of
  Mathematical Statistics Beachwood, Ohio, USA}, \bibinfo{year}{1983}).

\bibitem{HandbookIntegralEquations}
\bibinfo{author}{Polyanin, A.~D.} \& \bibinfo{author}{Manzhirov, A.~V.}
\newblock \emph{\bibinfo{title}{Handbook of integral equations}}
  (\bibinfo{publisher}{CRC press}, \bibinfo{year}{2008}).



\end{thebibliography}

\end{document}